\documentclass[11pt]{article}

\usepackage{amsmath,amssymb}
\usepackage{graphicx}
\usepackage{caption}
\usepackage{color}
\usepackage{dcolumn}
\usepackage{bm}
\usepackage{float}
\usepackage{hyperref} 
\usepackage{apacite}
\usepackage{mathtools}
\newcommand{\mathdash}{\relbar\mkern-9mu\relbar}
\usepackage[round]{natbib}
\usepackage{soul}
\usepackage{fullpage}

\newcommand{\indep}{\perp \!\!\! \perp}
\newcommand{\argmin}{\operatornamewithlimits{argmin}}

\title{Differential Network Analysis: A Statistical Perspective\thanks{To appear as an Advanced Review in \emph{WIREs Computational Statistics}.}}

\author{Ali Shojaie \\ Department of Biostatistics, University of Washington, Seattle WA}

\begin{document}
\maketitle

\begin{abstract}
Networks effectively capture interactions among components of complex systems, and have thus become a mainstay in many scientific disciplines. Growing evidence, especially from biology, suggest that networks undergo changes over time, and in response to external stimuli. In biology and medicine, these changes have been found to be predictive of complex diseases. They have also been used to gain insight into mechanisms of disease initiation and progression. Primarily motivated by biological applications, this article provides a review of recent statistical machine learning methods for inferring networks and identifying changes in their structures. 
\end{abstract}

\section*{\sffamily \Large INTRODUCTION} 

Networks are ubiquitous in many scientific disciplines. They are widely used to capture interactions among components of complex systems, and to glean insight into how these interactions shape the system's behavior. 
The latter is often achieved by comparing networks over time and/or in different states, a task referred to as \emph{differential network analysis} \citep{ideker2012}.

Differential network analysis has become particularly popular in biological studies, where growing evidence suggests that interactions among components of biological systems can vastly change over (evolutionary) time \citep{borneman2007, schmidt2010}, when the system responds to external stimuli \citep{bar2004, luscombe2004}, or in disease conditions \citep{hussain2006, goh2007}. For instance, changes in gene, protein and metabolite networks have been found to be associated with the onset and progression of various diseases \citep{zhong2009, zhang2016, west2012, ma2019diff}. Similarly, changes in brain connectivity networks have been successfully used as predictive biomarkers for neurodegenerative diseases \citep{chuang2007, taylor2009}. 

Let $G = (V,E)$ be a network with nodes $V = \{1, 2, \ldots, m\}$ and edge set $E \subseteq V \times V$. Changes in $G$ can be due to changes in its nodes, $V$, its edges, $E$, or both. Changes in the node set are common in social and communication networks, where both $V$ and $E$ can change as the network grows over time. In these settings, network edges---e.g., social interactions or internet connections---are directly observed and the primary goal is to understand the mechanisms of network growth \citep{durrett2007}. In contrast, in this paper we focus on the setting where the node set $V$ is fixed and the goal is to identify changes in network edges, $E$. Identifying such changes is of primary interest in the study of biological systems, where network nodes---e.g., genes or brain regions---can be measured, but network edges are often not directly observed. 
In fact, despite recent progress in developing assays for identifying interactions among genes and proteins \citep{stelzl2005, krogan2006, tarassov2008}, and changes in interactions in different biological conditions \citep{barrios2005}, interactions in biological systems and changes in those interactions are commonly inferred from measurements on the nodes. 
Primarily motivated by the challenges in biological applications, this paper reviews statistical methods for identifying changes in the edge set, $E$, inferred from $n$ observations on each node $j \in V$. To this end, we first briefly review probabilistic graphical models \citep{lauritzen1996}, which are the primary building blocks for inferring network edges. We then review statistical methods for differential network analysis. 

Throughout the paper, random variables are denoted by capital letters (e.g., $X$ and $X_j$), scalar parameters are denoted by lower case Greek letters (e.g., $\theta$) and parameter vectors/matrices are denoted by uppercase Greek letters (e.g., $\Theta$). Matrices of observations are denoted by Calligraphic letters (e.g., $\mathcal{X}$) and single observations are denoted by the corresponding lower case letters (e.g., $x_{ij}$). 

\section*{\sffamily \Large Background: Learning Network Structures}
Probabilistic graphical models are widely used to summarize dependency relationships among random variables \citep{lauritzen1996}, and to learn such dependencies from observations on the variables \citep{drton2017}. For a graph $G = (V,E)$, the set of nodes $V = \{1, \ldots, m\}$ is associated with random variables $X_1, \ldots, X_m$, and the edge set $E$ captures dependency relationships among the variables. The edges in $E$ can be \emph{directed} or \emph{undirected}. 


Directed graphical models are often used to capture causal relationships among random variables, with a directed edge $j \to k$ representing a direct causal effect of $X_j$ on $X_k$. The special case of directed acyclic graphs (DAGs)---where there are no directed cycles in $G$---corresponds to well-known \emph{Bayesian networks} \citep{pearl2009}, which have found many applications in biological \citep{markowetz2007} and social \citep{babin2012} sciences, as well as machine learning \citep{koller2009}. 

As expected, learning directed causal graphs from observational data is challenging and often impossible, or only possible under (uncheckable) identifiability assumptions \citep{peters2013}. This is because multiple DAGs may have the same likelihood and may thus be indistinguishable from data. Instead, the completed partially directed acyclic graph (CPDAG) representing the class of Markov equivalent DAGs is often estimated from observational data. Despite recent progress \citep{shojaie2010, wang2018, ghoshal2019, manzour2019}, existing methods for differential analysis of directed networks are in their infancy. As such, this review primarily focuses on differential analysis of undirected networks; references to recent work on differential analysis of directed networks are given in the Further Readings section. 

Methods for learning the structure of undirected networks can be broadly categorized into methods based on (i) \emph{marginal} and (ii) \emph{conditional} associations among variables, $X_1, \ldots, X_m$. These two classes of methods are reviewed in the remainder of this section. 


\subsection*{\sffamily \large Learning Networks from Marginal Associations}

Marginal inference procedures declare an (undirected) edge between two variables $X_j$ and $X_k$ if and only if they are dependent on each other. In practice, the dependence is often characterized by a marginal association measure, $\rho(X_j, X_k)$. In that case,  two nodes $j$ and $k$ are connected in $G$, i.e., $j \mathdash k \in E$, if and only if $\rho(X_j, X_k) \ne 0$. 

In the simplest case, the marginal association network is defined based on the (Pearson) correlation between $X_j$ and $X_k$. In practice, this simple approach, which is widely used in biological settings \citep{junker2008}, amounts to calculating the sample correlation coefficient between each pair of variables, $X_j$ and $X_k$, or, equivalently, the $(j,k)$ entry of the empirical correlation matrix of $X_1, \ldots, X_m$, denoted $S$. Learning the network structure then corresponds to selecting a subset of non-diagonal entries of $S$. This can be achieved by testing whether each $\rho(X_j, X_k)$ is zero, using, e.g., the Fisher's transformation of sample correlations \citep{fisher1921}, which can be used to test the hypothesis of no correlation, $H_0: \rho(X_j, X_k) = 0$.  
As an alternative, the network structure can be learned by identifying the set of correlations that are larger in magnitude than a pre-specified threshold $\kappa$. The threshold $\kappa$ plays the role of a \emph{tuning parameter}, and can be selected to achieve a certain level of sparsity in the network \citep{wang2006}, or to obtain a network that satisfies a certain degree distribution \citep{langfelder2008}. 

While simple, the Pearson correlation in the above procedure only captures linear dependencies. This is appropriate if $X_1, \ldots, X_m$ are jointly normally distributed. 
However, multivariate normality (or presence of linear dependencies \citep{khatri1976}) is a stringent assumption that may not hold in practice. As an alternative, rank-based correlation measures, such as Spearman correlation or Kendal's-$\tau$, or nonparametric measures of marginal association, such as mutual information \citep{margolin2006} or kernel-based measures of dependence \citep{yamanishi2004} can be used to test whether each pair of variables, $X_j$ and $X_k$, are independent. The network structure can then be learned by from $p$-values for testing independence among variables from each of these approaches, or by applying a pre-specified threshold. 

Regardless of the choice of association measure, the above network learning procedures have another limitation: marginal measures of associations cannot distinguish between direct and indirect relationships. As a simple example, consider three normally distributed variables $X_1, X_2$ and $X_3$. 
Suppose the true network $G$ consists of two edges, $1\mathdash 2$ and $1 \mathdash 3$. Further, suppose the true correlation between $X_1$ and both $X_2$ and $X_3$ is $0.8$. In other words, $\rho(X_1, X_2) = \rho(X_1, X_3) = 0.8$. But this implies that $\rho(X_2, X_3) = 0.64$! Thus, with enough observations, the network learned from the (correctly specified) marginal  association measure would incorrectly include the edge $2 \mathdash 3$. 

Despite its simplicity, the above example illustrates a major limitation of network inference based on marginal associations. Unfortunately, the same issue also arises with other distributions and other measures of associations. Network learning procedures based on conditional measures of associations, discussed next, try to address this limitation. 

\subsection*{\sffamily \large Learning Networks from Conditional Associations}

Undirected graphical models, also known as Markov random fields (MRF), represent conditional dependence relationships between a set of random variables. For random variables $\{X_1, X_2, \ldots, X_m\}$, an MRF is associated with an undirected graph $G=(V,E)$ with vertex set $V = \{1, 2, \ldots, m\}$ and undirected edges $E \subseteq V \times V$, such that the absence of an edge between nodes $j$ and $k$ indicates that $X_j$ and $X_k$ are conditionally independent given all other variables, i.e., $X_{\backslash \{j,k\}}$ \citep{lauritzen1996}. In the smallest such graph $G$, known as the \emph{conditional independence graph}, there is an edge between $j$ and $k$, i.e. $j \mathdash k \in E$, if and only if $X_j$ and $X_k$ are conditionally dependent given all other variables \citep{lauritzen1996}. 

Given $n$ observations from each random variable $X_j; \, j \in V$, learning the conditional independence graph (CIG) corresponds to identifying pairs of random variables that are independent given all other variables. While learning networks from conditional associations, and in particular the CIG, is more challenging than learning based on marginal associations, edges in a CIG capture unconfounded associations among variables and may thus be more scientifically meaningful. For instance, in the simple example of the previous section, the \emph{partial correlation} between $X_2$ and $X_3$ after adjusting for $X_1$ is indeed zero. Thus, the CIG correctly captures the association among variables. 

When $m = |V|$ is small compared to $n$, the CIG can be learned nonparametrically, using, e.g., nonparametric procedures for testing conditional independences, such as conditional mutual information \citep{margolin2006}, or kernel-based procedures \citep{yamanishi2004}. However, nonparametric procedures become computationally challenging, if not prohibitive, when $m$ is large. Moreover, it is not straightforward to extend such nonparametric procedures to high-dimensional settings, i.e., when $m > n$. In contrast, characterizing conditional independence is often easier if the family of probability distributions corresponding to $G$ is represented by finite-dimensional parameters. Such parametric models can also be more easily extended to high-dimensional settings. Finally, existing procedures for differential network analysis mainly consider parametric graphical models. Therefore, the rest of this section is primarily focused on parametric models, and we only provide a brief review of semi- and non-parametric graphical modeling approaches.

\subsubsection*{\sffamily \normalsize Gaussian Graphical Models}

The most well known, and most widely studied, example of probabilistic graphical models is the class of Gaussian graphical models (GGM), wherein $\{X_1, X_2, \ldots, X_m\}$ are jointly Gaussian. Formally, in a GGM, $(X_1, X_2, \ldots, X_m) \sim N(\mu, \Sigma)$, where $\mu \in \mathbb{R}^m$, $\Sigma \in \mathbb{S}^m_{+}$ and $\mathbb{S}^m_{+}$ denotes the set of symmetric positive definite matrices. In this case, any two variables $X_j$ and $X_k$ are conditionally independent, given all other variables $X_{\backslash \{j,k\}}$, if and only if the $(j,k)$ entry of the inverse covariance, or precision, matrix, $\Omega = \Sigma^{-1}$, is zero \citep{lauritzen1996}. Formally, 
\begin{equation}\label{eqn:GGM}
X_j \indep X_k \mid X_{\backslash \{j,k\}} \,\, \Leftrightarrow \,\, \Omega_{j,k} = \Omega_{k,j} = 0. 
\end{equation}

Equation~\eqref{eqn:GGM} implies that in the Gaussian case, the CIG is fully characterized by the precision matrix, $\Omega$. This characterization suggests the following simple estimation strategy: Let $\mathcal{X}$ be the $n \times m$ data matrix corresponding to $n$ i.i.d. observations for \emph{centered} variables $\{X_1, X_2, \ldots, X_m\}$ (so, $\sum_{i=1}^n{x_{ij}} = 0$). Then, calculate the empirical covariance matrix, $S = (n-1)^{-1} \mathcal{X}^\top \mathcal{X}$, and estimate the CIG based on nonzero entries of $S^{-1}$, by applying a threshold (similar to $\kappa$ discussed earlier for marginal association networks) or using an inference procedure \citep{drton2004}. 

While the above strategy is straightforward, the inverse of the empirical covariance matrix, $S^{-1}$, may not be well-conditioned even when $n > m$ \citep{dempster1972}. Moreover, the inverse does not even exist in the high-dimensional setting, where $m > n$. An alternative strategy is to directly calculate the partial correlations among pairs of variables, which are well known measures of conditional independence for Gaussian random variables. The partial correlation between $X_j$ and $X_k$ can be computed by first regressing each of them on the other variables, $X_{\backslash \{j,k\}}$, and then calculating the correlation between the residuals from these two regressions \citep{hair1998}. Partial correlations between $X_j$ and other variables can also be more directly obtained by regressing $X_j$ on all other variables. More specifically, suppose the variables are centered and scaled, and consider $m$ linear regressions
\begin{equation}\label{eqn:partialcor}
X_j = \sum_{k \ne j} \beta_{jk} X_k + \delta_j,	\quad j = 1, \ldots, m. 
\end{equation}
Then, $\beta_{jk}$ is the partial correlation between $X_j$ and $X_k$ given $X_{\backslash \{j,k\}}$. Moreover, $\beta_{jk} = - \Omega_{jk}/\Omega_{jj}$ \citep{meinshausen2006}. Thus, nonzero conditional independence relationships estimated based on $\Omega_{jk}$ and $\beta_{jk}$ coincide, leading to (asymptotically) equivalent estimates of the CIG. 

A potential drawback of regressions-based estimation of the CIG in \eqref{eqn:partialcor} is that, given a fixed sample size $n$, estimated conditional independences between $X_j$ and $X_k$ based on $\beta_{jk}$ and $\beta_{kj}$ may not coincide. Nonetheless, this regression-based strategy can be easily generalized to high-dimensional settings, by, e.g., utilizing a sparsity-inducing penalty such as the lasso \citep{tibshirani1996}. This approach, known as \emph{neighborhood selection}, was first considered in the seminal work of \citet{meinshausen2006}, who also established the consistency of the estimated CIG in high-dimensional sparse settings. In this approach, the `neighborhood' of each node $j \in V$ is defined as variables with non-zero coefficients in $m$ penalized regressions of the form 
\begin{equation}\label{eqn:ns}
\widehat\beta_{jk} = \argmin_{\beta_{jk}}\Big\| \mathcal{X}_j - \sum_{k \ne j} \beta_{jk} \mathcal{X}_k \Big\|_2^2 + \lambda \sum_{k \ne j} |\beta_{jk}|,	\quad j = 1, \ldots, m.  
\end{equation}
Here, the tuning parameter $\lambda$ controls the sparsity of the estimated neighborhoods, defined as $\widehat{\mathrm{ne}}_j = \left\{k: \widehat\beta_{jk} \ne 0 \right\}$. To mitigate the potential discrepancy between the neighborhoods (e.g., those estimated based on $\widehat\beta_{jk}$ and $\widehat\beta_{kj}$), the authors then propose constructing the CIG based on either the intersection or the union of the estimated neighborhoods. 

\newcommand{\tr}{\operatornamewithlimits{trace}}
\newcommand{\logdet}{\operatornamewithlimits{logdet}}

Sparsity inducing penalties can also be used to directly estimate the precision matrix, $\Omega$. In this approach, first considered by \citep{yuan2007, banerjee2008} and popularized by the efficient \emph{graphical lasso} algorithm \citep{friedman2008}, $\Omega$ is estimated by minimizing the $\ell_1$-penalized negative log likelihood 
\begin{equation}\label{eqn:glasso}
\widehat\Omega = \argmin_{\Omega \in \mathbb{S}_+^m}\big\{\tr(S\Omega) - \logdet(\Omega) + \lambda \| \Omega \|_1\big\}, 
\end{equation}
where, as before, $S$ is the empirical covariance matrix, and for a square matrix $M$, $\tr(M)$ and $\logdet(M)$ denote the sum of its diagonal entries and the logarithm of its determinant, respectively. In graphical modeling applications, the $\ell_1$ penalty $\| \Omega \|_1 = \sum_{j,k} |\Omega_{jk}|$ is often replaced by the sum of absolute values of the off-diagonal entries of $\Omega$, $\|\Omega \|_{1,\mathrm{off}} = \sum_{j\ne k} |\Omega_{jk}|$. 

Since their introductions, various authors have considered other penalties for both neighborhood selection and penalized likelihood estimation approaches, and have also investigated asymptotic properties of these estimators \citep{rothman2008}. A number of other approaches have also been proposed, including symmetric estimation of partial correlations \citep{peng2009, khare2015} as well as Bayesian estimation strategies \citep{wang2012}. More comprehensive reviews of the relevant papers can be found in the recent book on estimation of covariance matrices \citep{pourahmadi2013} and the review paper on structure learning in graphical models \citep{drton2017}. 

\subsubsection*{\sffamily \normalsize Graphical Models for Other Probability Distributions}

A key reason for the popularity of GGMs and the extensive recent work in this area is the convenient characterization of conditional independence relations for Gaussian random variables by the inverse covariance, or  precision, matrix. 
However, joint normality is a stringent assumption that may not be satisfied in many real data applications \citep{voorman2013}. In particular, GGMs are not appropriate when the observations are discrete (e.g., binary or Poisson), have heavy-tail distributions (e.g., exponential), or their support is a subset of the real line (e.g., non-negative).

The main challenge in estimating CIGs for other distributions is that unlike in the Gaussian case, conditional independence relations between pairs of variables are not necessarily characterized by a single parameter. Instead, conditional independence relations are more generally  characterized by the Hammersley-Clifford Theorem \citep{besag1975}, which states that a probability distribution $\mathcal{P}$ with a strictly positive density defines a Markov random field (MRF) over a graph $G$ \emph{if and only if} its density, $f$, can be factorized over complete subgraphs, or \emph{cliques}, of $G$. While elegant and general, this characterization does not necessarily lead to tractable algorithms for estimating CIGs given observations from $\{X_1, \ldots, X_m\}$. That is because one would need to search over all possible subsets of the variables to find the cliques that define the MRF. 

In the special case of GGMs, the Hammersley-Clifford Theorem is considerably simplified: In this case, it suffices to only consider  \emph{pairwise interactions} among variables, which is efficiently learned from the precision matrix of $\{X_1, \ldots, X_m\}$. Motivated by this property, graphical models for other distributions have also been defined based on pairwise interactions among variables. Denoting by $f_j(X_j)$ and $f_{jk}(X_j , X_k)$ the \emph{node and edge potentials}, respectively, the density $f(x)$ for such a \emph{pairwise MRF} is proportional to 
\begin{equation}\label{eqn:pMRF}
	\exp\left( \sum_{j=1}^m f_j(X_j) + 
		\frac{1}{2} \sum_{(j,k) \in E} f_{jk}(X_j, X_k) \right). 
\end{equation}
Importantly, \eqref{eqn:pMRF} implies that $f_{jk} = 0$ for $j \mathdash k \notin E$. Thus, the CIG can be estimated by identifying nonzero edge potentials. This characterization can be further simplified by parametrizing the edge potentials by, e.g., assuming 
\begin{equation}\label{eqn:pMRFpar}
	f_{jk}(X_j, X_k) = \theta_{jk}X_j X_k = \theta_{kj}X_k X_j, 
\end{equation}
for parameters $\theta_{jk} \in \mathbb{R}$. Let $\Theta \in \mathbb{S}^{m}$ be the matrix with zero diagonal entries and off diagonal entries equal to $\theta_{jk}$. Then, similar to GGMs, conditional independence relations for this family can be simply learned from the entries of $\Theta$: $j \mathdash k \in E$ if and only if $\theta_{jk} = \theta_{kj} = 0$ \citep{wainwright2008}. 

With the parametrization in \eqref{eqn:pMRFpar}, a key remaining challenge in estimating CIGs for exponential families is computing the normalizing constant to ensure that the distribution specified in \eqref{eqn:pMRF} is well defined. 
To overcome this challenge, \citet{yang2012} consider the case where conditional distributions for each node, given all other nodes, are  \emph{generalized linear models} (GLMs). More specifically, setting $f_j(X_j) = \theta_j X_j$, they consider \emph{conditionally-specified} graphical models, where node-conditional distributions are GLMs proportional to
\begin{equation}\label{eqn:GLMGM}
	\exp\left( \theta_j X_j + \sum_{k \in \mathrm{ne}(j)} \theta_{kj}X_k X_j + g(X_j)\right), 
\end{equation}
$\mathrm{ne}(j) = \{k: k\mathdash j \in E\}$ is the neighborhood of $j$ in $G$, and $g(\cdot)$ is a function that specifies different GLM distributions. 

\citet{yang2012} show that the conditionally-specified model \eqref{eqn:GLMGM} leads to a unique joint probability distribution of the form 
\begin{equation*}
	\exp\left( \sum_j \theta_j X_j + \sum_{(j,k) \in E} \theta_{kj}X_k X_j + \sum_j g(X_j) - h(\Theta) \right), 
\end{equation*}
where $h(\Theta)$ is the normalizing constant. 
Various GLM distributions are then obtained by considering different  functions $g(\cdot)$. For instance, $g(x) = -X_j^2/2$ corresponds to the  Gaussian distribution, while $g(x) = 0$ corresponds to the Bernoulli distribution \citep{ravikumar2010}. 
\citet{chen2014, yang2014} and \citet{cheng2017} further extend this approach to estimate CIGs from \emph{mixed data}, where node-conditional distributions are specified by multiple GLM distributions, for instance, binary, Poisson and Gaussian. 
 
A key advantage of the conditionally-specified model \eqref{eqn:GLMGM} is that it allows bypassing the computation of the normalizing constant, and facilitates computationally-efficient estimation of CIGs for a broad class of distributions. In fact, for GLMs, estimating the pairwise MRF amounts to solving $m$ GLM regressions---$m$ logistic regressions for binary data (similar to \citet{ravikumar2010}), and $m$ Poisson regressions for Poisson variables (similar to \citet{yang2013} and \citet{allen2013}). High-dimensional pairwise MRF for these and other distributions can then be estimated by augmenting the conditional negative log-likelihoods corresponding to \eqref{eqn:GLMGM} with a sparsity inducing penalty on $\Theta$, such as lasso. This approach is thus a natural extension of the neighborhood selection estimator of \citet{meinshausen2006} for other distributions in the exponential family. 

While computationally convenient, conditionally-specified models are not guaranteed to result in a symmetric network estimate (as discussed in the case of GGMs). To circumvent the latter shortcoming, few authors have proposed estimation strategies similar to conditionally-specified models that result in symmetric network estimates \citep[see, e.g.,][]{drton2017}. 
An alternative strategy for bypassing the computation of the normalizing constant, which can be used to directly obtain symmetric network estimates, is the \emph{score matching} approach of \citet{lin2016}. In this approach, the loss function is defined as the Fisher information distance between the gradients, with respect to observations $x$, of true and candidate log densities. 
Using integration-by-parts, \citet{hyvarinen2005} showed that under mild conditions, the empirical loss for a candidate density $f$ can be written as the average, over $n$ observations, of
\[
	\frac{1}{2} \| \nabla_x \log f(x) \|_2^2 + \Delta_x \log f(x),
\]
where $\nabla_x$ and $\Delta_x$ denote the gradient and Laplace operators with respect to $x$. 

\citet{lin2016} equipped the score matching loss with an $\ell_1$ penalty to obtain estimates of high-dimensional graphical models for distributions in the exponential family with absolutely continuous densities. Using the generalized score matching loss of \citet{hyvarinen2007}, they also extended this approach to distributions with densities supported over a subset of $\mathbb{R}$. See \citet{yu2018, yu2019} for further generalizations of this approach.

\subsubsection*{\sffamily \normalsize Semi-parametric and Nonparametric Graphical Models}

While computationally attractive and statistically efficient, parametric graphical models can lead to biased and incorrect CIG estimates if their underlying model does not hold. As an alternative to parametric models, few authors have recently considered semi- and non-parametric estimation of graphical models. Early work in this area considered the Gaussian copula or nonparanormal distribution \citep{liu2009, dobra2011}; instead of assuming multivariate normality, the nonparanormal model posits that for some (unknown) monotone functions $h_1, \ldots h_m$ the transformed variables $h_1(x_1), \ldots, h_m(x_m)$ have a multivariate normal distribution with mean zero and precision matrix $\Omega$. While estimating the unknown functions $h_j, j=1, \ldots, m$ seems difficult at first glance, \citet{liu2012} and \citet{xue2012} show that this approach is equivalent to estimating the CIG by plugging in a rank-based correlation matrix, such as Spearman correlation or Kendal's $\tau$ into the graphical lasso optimization problem \eqref{eqn:glasso}. 

The nonparanormal graphical model can be efficiently estimated and provides a natural generalization of the graphical lasso estimator. However, \citet{voorman2013} show that the nonparanormal model can be restrictive, and propose, as an alternative, conditionally-specified additive graphical models, by assuming 
\[
	X_j \mid X_{\backslash j} = \sum_{k \in \mathrm{ne}(j)} f_{jk}(X_k) + \varepsilon_j,  
\]
where $\varepsilon_j$ is a mean-zero noise variable. 
In this model, $X_j \indep X_k$ given other variables if and only if $f_{jk} = f_{kj} = 0$. Thus, in high dimensions, the CIG can be estimated by fitting $m$ penalized nonparametric regressions. \citet{voorman2013} consider a basis expansion  approach and use a joint standardized group lasso penalty \citep{simon2012standardization} to enforce both $f_{jk}$ and $f_{kj}$ to zero in order to estimate the neighborhood of each node in $G$. 
Other related ideas include the graphical random forest estimator of \citet{fellinghauer2013}, the kernel-based estimator of  \citet{lee2016}, as well as nonparametric approaches for exponential densities in \citet{sun2015learning} and \citet{suggala2017expxorcist}.

\section*{\sffamily \Large Statistical Methods for Differential Network Analysis}

Before reviewing recent developments in statistical methods for differential network analysis, we discuss relevant hypotheses and measures of difference between networks. 
For simplicity, we restrict the discussion to comparing two networks, $G^1$ and $G^2$ with the same node set $V$ and edges sets $E^1$ and $E^2$, or, equivalently, adjacency matrices $A^1$ and $A^2$.   
In general, $E^1$ and $E^2$ may have been directly observed, obtained from experiments, or learned from observations on the nodes via graphical modeling approaches. However, as mentioned in the Introduction, we focus primarily on networks inferred using graphical modeling methods. For instance, in the case of GGMs, $A^s, s \in \{1,2\}$ may correspond to estimated partial correlation matrices, $\widehat\Omega^s, s \in \{1,2\}$.

Various notions of difference between $A^1$ and $A^2$ can be considered. For instance, we may be interested in identifying \emph{global} differences between $A^1$ and $A^2$, i.e., whether $A^1 = A^2$. However, similar to testing for equality of vectors of parameters, different norms or distance measures can be used to assess whether $A^1$ and $A^2$ are the same. For instance, one can examine the difference between \emph{weighted} adjacency matrices, by examining the value of $\| A^1 - A^2 \|$ for some matrix norm. In the case of GGMs, this can be achieved by examining $\| \widehat\Omega^1 - \widehat\Omega^2 \|$. Alternatively, one can consider the \emph{structural Hamming distance} \citep{diestel2012} between $A^1$ and $A^2$, which counts the total number of edge differences between the two networks. Compared to the norm-based approach, which takes the \emph{quantitative} values of estimated parameters into account, this approach assesses \emph{qualitative} differences between the two networks. Finally, the topology of the space of networks offers additional measures of differences between $A^1$ and $A^2$, including (potentially vector-valued) summary measures of the two networks, such as the size and/or number of clusters, the average connectivity, or the degree distribution; see \citet{shojaie2017} for examples of such measures. 

In many applications, \emph{local} differences between the two networks, including differences in individual edges, neighborhoods or subnetworks, can also be of interest. This is especially the case in biological applications, where \emph{network-based biomarkers} can be used to interrogate mechanisms of diseases initiation and progression \citep{erler2010, gomez2014, liu2016}. Identifying local differences between networks can also be of interest following an affirmative global test of difference between the two networks. As in the case of global differences, local differences between two networks can be assessed qualitatively or quantitatively. For instance, in the case of GGMs, one may be interested in identifying node pairs $(j,k)$ such that $\widehat\Omega^1_{jk} \ne \widehat\Omega^2_{jk}$. Alternatively, instead of looking at quantitative differences between parameters, we may want to identify node pairs $(j,k)$ such that $j \mathdash k \in G^1$ but $j \mathdash k \notin G^2$. In the Gaussian case, such qualitative differences can be identified by comparing the zero/nonzero patterns of $\widehat\Omega^1$ and $\widehat\Omega^2$; for instance, by identifying node-pairs $(j,k)$ such that $\text{supp}\left(\widehat\Omega^1_{jk}\right) \ne \text{supp}\left(\widehat\Omega^2_{jk}\right)$, where $\text{supp}(\omega) = 1$ if $\omega \ne 0$ and $0$ otherwise. 

\begin{figure}[t]
\centering
\includegraphics[scale=0.8]{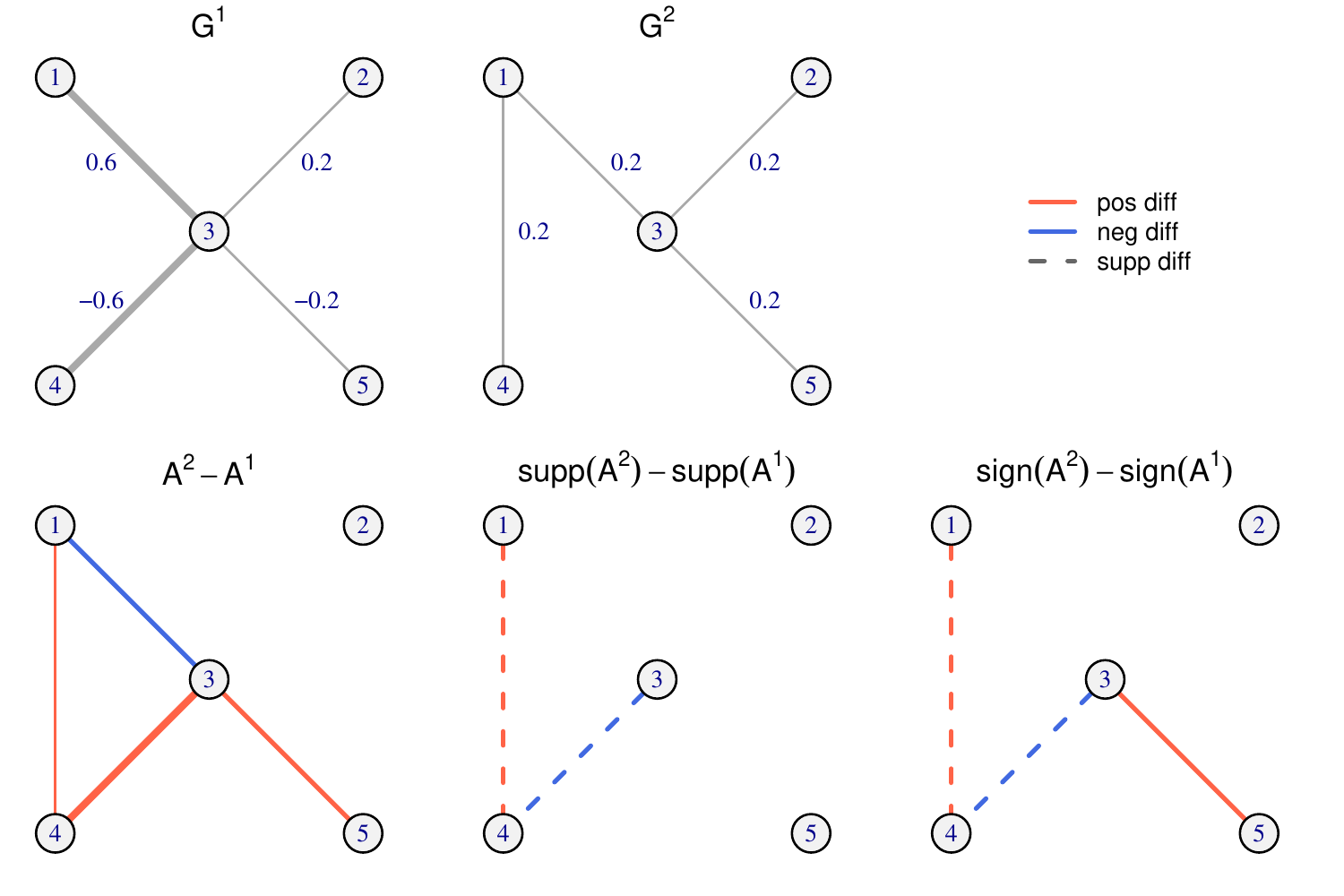}
\caption{Illustration of different notions of difference in networks. \textbf{Top}: Hypothetical networks for two populations; here, networks correspond to two GGMs and adjacency matrices $A^1$ and $A^2$ correspond to (true) partial correlations among nodes. \textbf{Bottom}: Differential networks based on differences in \emph{values} of adjacency matrices (left); differences in \emph{supports} of the adjacency matrices (center); and differences in \emph{signs} of adjacency matrices (right).}\label{demofig}
\end{figure}

Examples of quantitative and qualitative differences in networks are depicted in Figure~\ref{demofig}. This simple example highlights different insights and conclusions based on different notions of network difference: the differential network based on \emph{values} of partial correlations ($A^2 - A^1$, bottom-left) captures differences in signs and magnitudes of model parameters; the differential network based on \emph{supports} of $A^2$ and $A^1$ (bottom-center) captures differences in edge structures; and the differential network based on differences in \emph{signs} (bottom-right) captures both support and sign differences between. The choice of the appropriate notion of difference depends on the application. In particular, as discussed in the remainder of this section, qualitative methods/tests may better capture differences in the \emph{structures} of underlying networks, while quantitative methods could offer higher power for identifying differences in \emph{parameters} of graphical models used to learn the networks. 

In the following, we discuss existing statistical approaches that examine various notions of difference between two networks (global vs. local and qualitative vs. quantitative). Given the current state of the literature, we focus primarily on methods for Gaussian observations, and briefly review methods for other graphical models at the end.

\subsection*{\sffamily \large Global Tests of Network Differences}

Naturally, the global null hypothesis of no difference between two GGMs, i.e., $H_0: E^1 = E^2$, can be tested by examining whether correlation,  or partial correlation, matrices in the two populations are different. Formally, two GGMs are the same if $H_0: \Sigma^1 = \Sigma^2$, or, equivalently, $H_0: \Omega^1 = \Omega^2$, holds. However, as mentioned earlier, these matrix-based hypotheses can be tested using different matrix norms and summaries. Regardless of the choice of norm/summary, a key challenge arises from high-dimensionality: When $m \gg n$ classical estimates of $\Sigma^s, s\in\{1,2\}$ may be too noisy for an unbiased tests, and estimating $\Omega^s, s\in\{1,2\}$ requires regularization methods that rely on sparsity. 

Motivated by classical multivariate methods, early tests of difference between high-dimensional correlation matrices \citep{schott2007, li2012} were based on the Frobenius norm, $\|\Sigma^1 - \Sigma^2\|_F^2 = \sum_{j=1}^m\sum_{k=1}^m{\left(\Sigma^1_{jk} - \Sigma^2_{jk}\right)^2}$. These tests are sensitive to orchestrated weak changes in entries of the correlation matrices, but may have low power if few correlations are significantly different, but the majority are similar. In contrast, methods based on maximum entries of matrices \citep{cai2016, chang2017} are sensitive to large differences between individual correlations, i.e., sparse but large differences. Other approaches have utilized eigen-structures \citep{srivastava2010} and random matrix projections \citep{wu2015}. In a recent work, \citet{zhu2017} proposed a test based on sparse leading eigenvectors that can detect both sparse and weak differences. 

A potential advantage of the above methods for testing differences in covariance matrices is that they can also be applied to pre-specified subsets of nodes. More specifically, for a subset $U \subseteq V$ of nodes, the above methods can test $H_0: \Sigma^1_{U,U} = \Sigma^2_{U,U}$. 
Such tests are particularly relevant in \emph{pathway enrichment analysis} \citep{khatri2012}, where $U$ is the set of nodes corresponding to a biological pathway, and the goal is to determine whether the distributions of random variables $X_j$ for $j \in U$ are the same across two populations. Similar problems also arise in other applications, for instance, when interrogating composite brain regions \citep{tryputsen2015}. Both nonparametric methods, such as the energy statistic \citep{szekely2013}, and permutation-based approaches \citep{subramanian2005, tian2005} have been used to test for differences in distributions. However, more recent approaches have focused on accounting for the topology of the underlying networks \citep{khatri2012} by utilizing the full power of graphical models. 
For instance, assuming normality, the topologyGSA method \citep{massa2010} first tests for equality of covariance matrices, $\Sigma^1 = \Sigma^2$. Depending on the outcome of this test, pathway enrichment is determined by testing for differences in means, i.e. $\mu^1 = \mu^2$: if equality of covariances is not rejected, a multivariate analysis of variance (MANOVA) \citep{smith1962} is used, whereas the Behrens-Fisher method \citep{anderson1958} is used if covariances are found to be different. 
Similarly, DEGraph \citep{jacob2012} also starts with testing $\Sigma^1 = \Sigma^2$. If this hypothesis is rejected, then the pathway is declared to be enriched. If not, differences in means are tested using a Hotelling's $T^2$ statistic \citep{hotelling1931} using the pooled estimate of the covariance matrix. 
The NetGSA framework \citep{shojaie2009, shojaie2010, ma2016netgsa} is also related, but takes a different perspective; it  combines differences in mean and covariance matrices between the two populations by considering a latent variable model, and defines a contrast vector based on covariances. Aside from details of testing procedures, another key difference between NetGSA and other methods is that it uses the observations in each population to learn/update the estimated network in each condition, and thus accounts for differential connectivity in the two networks. See \citet{ma2019} for more discussions and a recent review of topology-based pathway enrichment methods. 

\begin{figure}[t]
\centering
\includegraphics[width=0.9\textwidth]{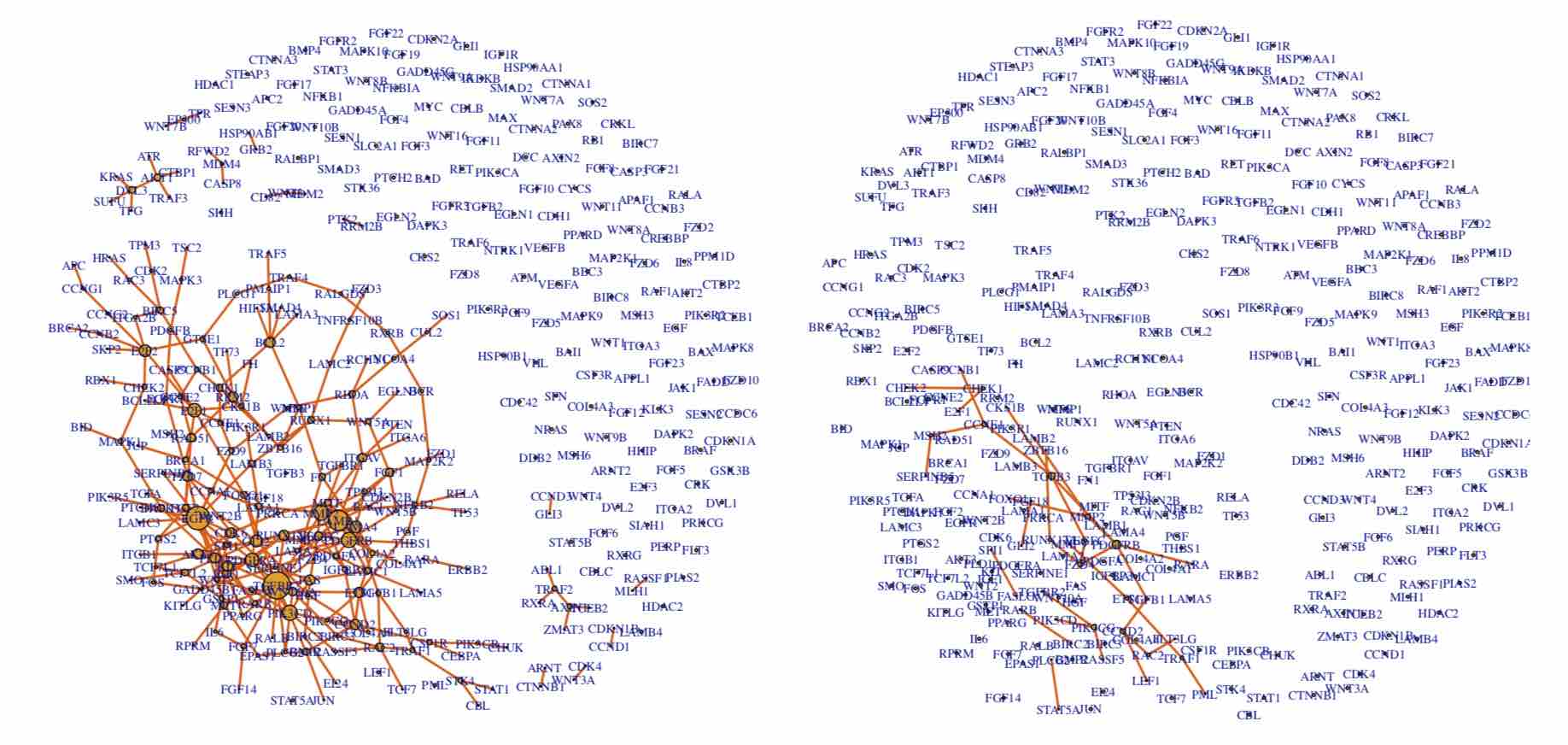}
\caption{Differential network analysis in subtypes of breast cancer. The two networks show edges identified as significant in only one breast cancer subtypes (\textbf{Left}: ER+; \textbf{Right}: ER-). They correspond to interactions among a subset of $m=358$ cancer-related genes, and are inferred using gene expression measurements from the Cancer Genome Atlas (TCGA).}\label{netfig}
\end{figure}

\subsection*{\sffamily \large Estimating Multiple GGMs and Their Differences}
Biological systems are inherently robust \citep{Kitano2004}. Therefore, despite potential differences, networks in similar conditions or populations are expected to share many common edges. For instance, gene regulatory networks in different cancer subtypes, for instance ER+ and ER- subtypes of breast cancer in Figure~\ref{netfig}, are expected to share many edges. It therefore makes sense to account for these common edges. This is particularly the case when estimating high-dimensional graphical models, where the small sample size, compared to the number of variables/features, is a key challenge. 
Recent graphical modeling approaches that try to account for common edges in networks in order to better delineate their differences can be broadly categorized into two classes: \emph{joint estimation of multiple graphical models} and \emph{direct estimation of differences between graphical models}. 

In joint estimation of multiple graphical models, the goal is to borrow information across populations/conditions in order to better estimate the networks in each condition. For instance, when estimating two GGMs, this can be achieved by encouraging the entries of the precision matrices to be similar to each other. More specifically, let $\Omega^1_{jk}$ and $\Omega^2_{jk}$ be $(j,k)$ entries of precision matrices in two populations. Then, joint estimation strategies encourage the estimates of $\Omega^1_{jk}$ and $\Omega^2_{jk}$ to be similar to each other. To achieve this goal, \citet{guo2011} proposed to re-parametrize the entries of the precision matrices as the product of a common parameter (for both populations) and a population-specific parameter. Formally, for $s \in \{1,2\}$, they let $\Omega^s_{jk} = \Theta_{jk} \Gamma^s_{jk}$, where to avoid sign ambiguity, $\Theta_{jk}$ is restricted to be nonnegative. The graphical models are then jointly estimated by replacing the $\ell_1$ penalty in the graphical lasso problem \eqref{eqn:glasso} with two penalties on $\Theta_{jk}$ and $\Gamma^s_{jk}:$ 
\begin{equation*}
	\lambda_1 \sum_{j \ne k} \Theta_{jk} + \lambda_2 \sum_s \sum_{j \ne k} | \Gamma^s_{jk}|.
\end{equation*}
The first penalty encourages sparsity in both $\Omega^1_{jk}$ and $\Omega^2_{jk}$, and hence improves the selection of common zero coefficients in the precision matrices. If $\Theta_{jk} \ne 0$, then the second penalty induces condition-specific sparsity in each of the precision matrices.  

The proposal of \citet{guo2011} leads to a non-convex optimization problem, and potential challenges in large-scale networks. As an alternative, \citet{danaher2014} proposed to directly augment the graphical lasso problem \eqref{eqn:glasso} with a second penalty to encourage similarity among $\Omega^s_{jk}, s \in \{1,2\}$ coefficients. In particular, they proposed two penalties: a \emph{group lasso} penalty \citep{yuan2006}, $\sum_{jk}\sqrt{\left(\Omega^1_{jk}\right)^2 + \left(\Omega^2_{jk}\right)^2}$, and a \emph{fused lasso} penalty \citep{tibshirani2005}, $\sum_{jk} | \Omega^1_{jk} - \Omega^2_{jk}|$. The group lasso penalty encourages similar sparsity patterns across the two populations, whereas the fused lasso penalty encourages the coefficients across the two populations to be equal to each other. 

While effective for jointly learning two networks, the strategies described above may not work well for learning \emph{multiple} GGMs. This is because they inherently assume that the networks in multiple (sub)populations are equally similar to each other. 
Addressing this shortcoming is the primary focus of a number of recent papers, including \citet{zhu2014, peterson2015, ma2016, saegusa2016}. To achieve this goal, \citet{zhu2014} and \citet{ma2016} generalize the fused and group lasso penalties, respectively, to account for the \emph{known} similarity structure among multiple networks. The methods by \citet{peterson2015} and \citet{saegusa2016} focus instead on the setting where the similarity structure is \emph{unknown}. In particular, \citet{peterson2015} propose a Bayesian approach by using a Markov random field (MRF) prior to learn the precision matrices in a mixture of Gaussian distributions. 
To overcome the computational challenges of Bayesian estimation of GGMs, this approach assumes that network edges are formed independently. \citet{saegusa2016} instead propose to use a Laplacian shrinkage penalty \citep{huang2011} based on a similarity structure learned from data. More specifically, instead of a fused or group lasso penalty, the authors propose to use $\sum_{j,k}\left[\sum_{s,s'} \pi_{s,s'} \left( \Omega^{s}_{jk} - \Omega^{s'}_{jk} \right)^2\right]^{1/2}$, where the \emph{data-driven weights} $\pi_{s,s'}$ capture the similarity among (sub)populations $s$ and $s'$. To justify this data-driven penalty, the authors establish the consistency of hierarchical clustering in high-dimensional settings and use the resulting clustering to define the similarity structure among (sub)populations. The idea of combining clustering and estimation of multiple graphical models was also considered in \citep{hao2017}, wherein clustering and graphical model estimation are combined into a single problem, which is solved using an Expectation Conditional Maximization (ECM) algorithm. 

Methods for joint estimation of multiple graphical models provide valuable insight into commonalities and differences between networks in different populations. However, when the primary scientific focus is on differences between networks, learning their common structures may be unnecessary and inefficient. As an alternative, \citet{zhao2014} proposed to directly estimate the difference of two GGMs. More specifically, the authors utilize the CLIME estimation framework \citep{cai2011} to estimate the sparse difference of two precision matrices, $\Delta = \Omega^2 - \Omega^1$ subject to a constraint motivated by the observation that the true covariance and precision matrices must satisfy
\[
	\Sigma^1 \Delta \Sigma^2 - (\Sigma^2 - \Sigma^1) = 0.
\]
The key advantage of this approach is it only assumes that the difference of the precision matrices, $\Delta$, is sparse, and not each of the precision matrices. However, solving the optimization problem for direct estimation of differences introduces additional challenges. To overcome these, the authors also propose an alternative formulation based on neighborhood selection. \citet{yuan2017} have recently proposed a more computationally-appealing alternative based on the D-trace loss \citep{zhang2014}, which is a special case of the score matching loss \citep{lin2016} discussed earlier; see also \citet{na2019estimating} for a related approach to learn differences in networks with latent (hidden) nodes. 

\subsection*{\sffamily \large Testing for Differences in Network Edges}

Unlike global tests of network differences, methods for joint estimation of multiple graphical models and their differences do not provide measures of uncertainty, such as confidence intervals and p-values. Thus, although they provide powerful tools for exploratory analysis and hypothesis generation, the methods discussed in the previous section have limited utility in scientific applications. In contrast, recent hypothesis testing procedures for single precision matrices \citep{Renetal2015, JankovavadeGeer2015, JankovavandeGeer2017, XiaLi2017} offer confidence intervals for entries of each precision matrix, $\Omega^s$,  and/or p-values for the null hypothesis $H_{0}: \Omega^s_{jk}=0$ for $j\neq k$. \citet{yu2019simultaneous} have further generalized this idea for inference in non-Gaussian graphical models using the framework fo generalized score matching \citep{yu2019}. 

Equipped with a multiple comparison adjustment procedure \citep[e.g.,][]{benjamini1995}, the above inference methods can be used to (asymptotically) control the probability of falsely detecting nonexistent network edges in each (sub)population. However, these inference procedures are not guaranteed to control the probability of false positives when testing differences between networks. To see this, consider testing the difference between the $(j,k)$ entry in two precision matrices, i.e., $\Omega^1_{jk}$ and $\Omega^2_{jk}$. Suppose we obtain confidence intervals for these parameters, using, e.g., the method of \citet{JankovavadeGeer2015}. These confidence intervals can be used to test the difference in \emph{support} of the two networks with respect to the $j \mathdash k$ edge, as illustrated in Figure~\ref{demofig}. (The confidence intervals can also be used to test for differences in \emph{signs} and \emph{values} of the precision matrices, but, for simplicity, here we focus only on the support.) If both confidence intervals cover zero, or if both do not overlap with zero, then we conclude, with high confidence, that the two networks are \emph{not} differentially connected at this edge. However, things become complicated if one confidence interval covers zero and the other does not. In this case, the optimistic conclusion is that the difference in coverage of confidence intervals points to differential connectivity between the two networks. However, that is not necessarily the case! The fact that one of the confidence intervals covers zero may simply be due to the low power of the inference procedure, especially if the true parameter or the sample size is small. This simple example highlights the primary limitation of single network inference for inferring differential connectivity between networks. 

Inference procedures for detecting differences in two GGMs directly examine whether the entries in the two precision matrices are equal. 
For instance, \citet{Xiaetal2015} tests whether $\Omega^1_{jk}=\Omega^2_{jk}$ using the connection between the entries of the precision matrix and the regression coefficients obtained from neighborhood selection \citep{meinshausen2006}. Alternatively, \citet{he2019} test the same hypothesis directly based on estimates of precision matrices using graphical lasso \citep{friedman2008}. 
As yet another alternative, \citet{Belilovskyetal2016} propose a test by directly estimating the difference between two vectors of regression coefficients using a multi-task fused lasso penalty. This approach offers an efficient framework for testing the difference between partial correlations, which may be of interest in some applications. 

As an alternative perspective to the above procedures, \citet{zhao2019} have recently argued that quantitative tests for differential analysis of undirected networks, e.g., tests based on differences between entries of precision matrices (or partial correlations) may not be desirable. In making this argument, they first point out that while GGMs are used for network inference, differences in parameter values (e.g. differences in partial correlations) may not be scientifically meaningful. Rather, the scientists are often interested in whether connectivity patterns are different. They also point out that because of their complex dependence patterns, GGM parameters corresponding to other edges may change if few edges in the network are rewired. As a result, tests based on \emph{quantitative} differences between GGM parameters could result in uncontrollable false positives if the goal is to identify differences in network structures. To circumvent these issues, \citet{zhao2019} propose a new framework, termed \emph{differential connectivity analysis} (DCA), for testing \emph{qualitative} differences in patterns of connectivity between two GGMs. However, testing qualitative hypotheses is more challenging and DCA requires additional assumptions.

\section*{\sffamily \Large CONCLUSIONS}
Differential network analysis is a promising new field with diverse biological applications \citep{sas2018, gambardella2013, ma2014, cabusora2005, troy2016}. Given that networks are often not directly observed in biological settings, statistical methods for identifying differences between networks will continue to be essential tools in this area. With few exceptions (see Further Readings), existing statistical and computational approaches have thus far primarily focused on undirected Gaussian graphical models (GGMs). Differential network analysis for non-Gaussian data and directed networks offer fruitful  opportunities of future research. Addressing the limitations of quantitative tests of differences between networks, discussed in \citet{zhao2019} and briefly reviewed in the previous section, would also be an important direction of future research. 

In addition to inferring networks based on activities of components of biological systems, a number of experimental platforms, such as ChIP-Seq and ChIP-chip assays \citep{landt2012}, have also been developed to interrogate the interactions among these components as well as changes in these interactions \citep{barrios2005}. These emerging assays offer the opportunity to more directly observe network edges or changes in the network structures. They may also be able to validate the findings from statistical/computational approaches, which is currently a key challenge. Designing efficient experiments based on these new assays \citep{kerr2001} and accounting, and adjusting for batch effects \citep{leek2010} are challenging but impactful areas of future research.

\section*{\sffamily \Large ACKNOWLEDGEMENTS}
I would like to thank three anonymous reviewers for their constructive feedback. I also thank Dr. Jing Ma for helpful input on an earlier version of the manuscript. 

\section*{\sffamily \Large FUNDING INFORMATION}
This work was made possible by the NSF grant DMS-1561814. Additional support from NIH grant R01GM114029 is also gratefully acknowledged. 

\subsection*{\sffamily \Large FURTHER READING}
Recent developments on statistical approaches for differential network analysis have started to focus on directed networks, and, in particular, directed acyclic graphs (DAGs) \citep{wang2018, ghoshal2019}, as well as graphical models for other data types \citep{cai2018, kim2019, zhao2019, he2019, yu2019simultaneous}. A number of software tools have also been developed that provide tests of differential connectivity based on permutation approaches \citep{Gilletal2014}, or by considering differences in marginal associations based on correlations, instead of conditional dependencies \citep{fukushima2013, mckenzie2016}. While these tools may not have strong theoretical support, or may test different hypotheses, they provide more convenient user interfaces and may be more computationally amenable for analysis of large networks.

\bibliographystyle{abbrvnat}
\bibliography{refs}

\begin{thebibliography}{137}
\providecommand{\natexlab}[1]{#1}
\providecommand{\url}[1]{\texttt{#1}}
\expandafter\ifx\csname urlstyle\endcsname\relax
  \providecommand{\doi}[1]{doi: #1}\else
  \providecommand{\doi}{doi: \begingroup \urlstyle{rm}\Url}\fi

\bibitem[Allen and Liu(2013)]{allen2013}
G.~I. Allen and Z.~Liu.
\newblock A local poisson graphical model for inferring networks from
  sequencing data.
\newblock \emph{IEEE transactions on nanobioscience}, 12\penalty0 (3):\penalty0
  189--198, 2013.

\bibitem[Anderson(2003)]{anderson1958}
T.~W. Anderson.
\newblock An introduction to multivariate statistical analysis (3rd edition),
  2003.

\bibitem[Babin and Svensson(2012)]{babin2012}
B.~J. Babin and G.~Svensson.
\newblock Structural equation modeling in social science research: Issues of
  validity and reliability in the research process.
\newblock \emph{European Business Review}, 24\penalty0 (4):\penalty0 320--330,
  2012.

\bibitem[Banerjee et~al.(2008)Banerjee, Ghaoui, and
  d’Aspremont]{banerjee2008}
O.~Banerjee, L.~E. Ghaoui, and A.~d’Aspremont.
\newblock Model selection through sparse maximum likelihood estimation for
  multivariate gaussian or binary data.
\newblock \emph{Journal of Machine learning research}, 9\penalty0
  (Mar):\penalty0 485--516, 2008.

\bibitem[Bar-Yam and Epstein(2004)]{bar2004}
Y.~Bar-Yam and I.~R. Epstein.
\newblock Response of complex networks to stimuli.
\newblock \emph{Proceedings of the National Academy of Sciences}, 101\penalty0
  (13):\penalty0 4341--4345, 2004.

\bibitem[Barrios-Rodiles et~al.(2005)Barrios-Rodiles, Brown, Ozdamar, Bose,
  Liu, Donovan, Shinjo, Liu, Dembowy, Taylor, et~al.]{barrios2005}
M.~Barrios-Rodiles, K.~R. Brown, B.~Ozdamar, R.~Bose, Z.~Liu, R.~S. Donovan,
  F.~Shinjo, Y.~Liu, J.~Dembowy, I.~W. Taylor, et~al.
\newblock High-throughput mapping of a dynamic signaling network in mammalian
  cells.
\newblock \emph{Science}, 307\penalty0 (5715):\penalty0 1621--1625, 2005.

\bibitem[Belilovsky et~al.(2016)Belilovsky, Varoquaux, and
  Blaschko]{Belilovskyetal2016}
E.~Belilovsky, G.~Varoquaux, and M.~B. Blaschko.
\newblock Testing for differences in gaussian graphical models: Applications to
  brain connectivity.
\newblock In D.~D. Lee, M.~Sugiyama, U.~V. Luxberg, I.~Guyon, and R.~Garnett,
  editors, \emph{Advances in Neural Information Processing Systems}, volume~29,
  pages 595--603. Curran Associates, Inc., Red Hook, NY, 2016.

\bibitem[Benjamini and Hochberg(1995)]{benjamini1995}
Y.~Benjamini and Y.~Hochberg.
\newblock Controlling the false discovery rate: a practical and powerful
  approach to multiple testing.
\newblock \emph{Journal of the Royal statistical society: series B
  (Methodological)}, 57\penalty0 (1):\penalty0 289--300, 1995.

\bibitem[Besag(1975)]{besag1975}
J.~Besag.
\newblock Statistical analysis of non-lattice data.
\newblock \emph{Journal of the Royal Statistical Society: Series D (The
  Statistician)}, 24\penalty0 (3):\penalty0 179--195, 1975.

\bibitem[Borneman et~al.(2007)Borneman, Gianoulis, Zhang, Yu, Rozowsky,
  Seringhaus, Wang, Gerstein, and Snyder]{borneman2007}
A.~R. Borneman, T.~A. Gianoulis, Z.~D. Zhang, H.~Yu, J.~Rozowsky, M.~R.
  Seringhaus, L.~Y. Wang, M.~Gerstein, and M.~Snyder.
\newblock Divergence of transcription factor binding sites across related yeast
  species.
\newblock \emph{Science}, 317\penalty0 (5839):\penalty0 815--819, 2007.

\bibitem[Cabusora et~al.(2005)Cabusora, Sutton, Fulmer, and
  Forst]{cabusora2005}
L.~Cabusora, E.~Sutton, A.~Fulmer, and C.~V. Forst.
\newblock Differential network expression during drug and stress response.
\newblock \emph{Bioinformatics}, 21\penalty0 (12):\penalty0 2898--2905, 2005.

\bibitem[Cai et~al.(2011)Cai, Liu, and Luo]{cai2011}
T.~Cai, W.~Liu, and X.~Luo.
\newblock A constrained ℓ 1 minimization approach to sparse precision matrix
  estimation.
\newblock \emph{Journal of the American Statistical Association}, 106\penalty0
  (494):\penalty0 594--607, 2011.

\bibitem[Cai et~al.(2018)Cai, Li, Ma, and Xia]{cai2018}
T.~Cai, H.~Li, J.~Ma, and Y.~Xia.
\newblock Differential markov random field analysis with an application to
  detecting differential microbial community networks.
\newblock \emph{Biometrika}, 103\penalty0 (1):\penalty0 1--16, 2018.

\bibitem[Cai and Zhang(2016)]{cai2016}
T.~T. Cai and A.~Zhang.
\newblock Inference for high-dimensional differential correlation matrices.
\newblock \emph{Journal of multivariate analysis}, 143:\penalty0 107--126,
  2016.

\bibitem[Chang et~al.(2017)Chang, Zhou, Zhou, and Wang]{chang2017}
J.~Chang, W.~Zhou, W.-X. Zhou, and L.~Wang.
\newblock Comparing large covariance matrices under weak conditions on the
  dependence structure and its application to gene clustering.
\newblock \emph{Biometrics}, 73\penalty0 (1):\penalty0 31--41, 2017.

\bibitem[Chen et~al.(2014)Chen, Witten, and Shojaie]{chen2014}
S.~Chen, D.~M. Witten, and A.~Shojaie.
\newblock Selection and estimation for mixed graphical models.
\newblock \emph{Biometrika}, 102\penalty0 (1):\penalty0 47--64, 2014.

\bibitem[Cheng et~al.(2017)Cheng, Li, Levina, and Zhu]{cheng2017}
J.~Cheng, T.~Li, E.~Levina, and J.~Zhu.
\newblock High-dimensional mixed graphical models.
\newblock \emph{Journal of Computational and Graphical Statistics}, 26\penalty0
  (2):\penalty0 367--378, 2017.

\bibitem[Chuang et~al.(2007)Chuang, Lee, Liu, Lee, and Ideker]{chuang2007}
H.-Y. Chuang, E.~Lee, Y.-T. Liu, D.~Lee, and T.~Ideker.
\newblock Network-based classification of breast cancer metastasis.
\newblock \emph{Molecular systems biology}, 3\penalty0 (1), 2007.

\bibitem[Danaher et~al.(2014)Danaher, Wang, and Witten]{danaher2014}
P.~Danaher, P.~Wang, and D.~M. Witten.
\newblock The joint graphical lasso for inverse covariance estimation across
  multiple classes.
\newblock \emph{Journal of the Royal Statistical Society: Series B (Statistical
  Methodology)}, 76\penalty0 (2):\penalty0 373--397, 2014.

\bibitem[Dempster(1972)]{dempster1972}
A.~P. Dempster.
\newblock Covariance selection.
\newblock \emph{Biometrics}, pages 157--175, 1972.

\bibitem[Diestel(2012)]{diestel2012}
R.~Diestel.
\newblock \emph{Graph theory: Springer graduate text gtm 173}, volume 173.
\newblock Reinhard Diestel, 2012.

\bibitem[Dobra et~al.(2011)Dobra, Lenkoski, et~al.]{dobra2011}
A.~Dobra, A.~Lenkoski, et~al.
\newblock Copula gaussian graphical models and their application to modeling
  functional disability data.
\newblock \emph{The Annals of Applied Statistics}, 5\penalty0 (2A):\penalty0
  969--993, 2011.

\bibitem[Drton and Maathuis(2017)]{drton2017}
M.~Drton and M.~H. Maathuis.
\newblock Structure learning in graphical modeling.
\newblock \emph{Annual Review of Statistics and Its Application}, 4:\penalty0
  365--393, 2017.

\bibitem[Drton and Perlman(2004)]{drton2004}
M.~Drton and M.~D. Perlman.
\newblock Model selection for gaussian concentration graphs.
\newblock \emph{Biometrika}, 91\penalty0 (3):\penalty0 591--602, 2004.

\bibitem[Durrett(2007)]{durrett2007}
R.~Durrett.
\newblock \emph{Random graph dynamics}, volume 200.
\newblock Cambridge university press Cambridge, 2007.

\bibitem[Erler and Linding(2010)]{erler2010}
J.~T. Erler and R.~Linding.
\newblock Network-based drugs and biomarkers.
\newblock \emph{The Journal of Pathology: A Journal of the Pathological Society
  of Great Britain and Ireland}, 220\penalty0 (2):\penalty0 290--296, 2010.

\bibitem[Fellinghauer et~al.(2013)Fellinghauer, B{\"u}hlmann, Ryffel,
  Von~Rhein, and Reinhardt]{fellinghauer2013}
B.~Fellinghauer, P.~B{\"u}hlmann, M.~Ryffel, M.~Von~Rhein, and J.~D. Reinhardt.
\newblock Stable graphical model estimation with random forests for discrete,
  continuous, and mixed variables.
\newblock \emph{Computational Statistics \& Data Analysis}, 64:\penalty0
  132--152, 2013.

\bibitem[Fisher(1921)]{fisher1921}
R.~A. Fisher.
\newblock On the `probable error' of a coefficient of correlation deduced from
  a small sample.
\newblock \emph{Metron}, 1:\penalty0 1--32, 1921.

\bibitem[Friedman et~al.(2008)Friedman, Hastie, and Tibshirani]{friedman2008}
J.~Friedman, T.~Hastie, and R.~Tibshirani.
\newblock Sparse inverse covariance estimation with the graphical lasso.
\newblock \emph{Biostatistics}, 9\penalty0 (3):\penalty0 432--441, 2008.

\bibitem[Fukushima(2013)]{fukushima2013}
A.~Fukushima.
\newblock Diffcorr: an r package to analyze and visualize differential
  correlations in biological networks.
\newblock \emph{Gene}, 518\penalty0 (1):\penalty0 209--214, 2013.

\bibitem[Gambardella et~al.(2013)Gambardella, Moretti, De~Cegli, Cardone,
  Peron, and Di~Bernardo]{gambardella2013}
G.~Gambardella, M.~N. Moretti, R.~De~Cegli, L.~Cardone, A.~Peron, and
  D.~Di~Bernardo.
\newblock Differential network analysis for the identification of
  condition-specific pathway activity and regulation.
\newblock \emph{Bioinformatics}, 29\penalty0 (14):\penalty0 1776--1785, 2013.

\bibitem[Ghoshal and Honorio(2019)]{ghoshal2019}
A.~Ghoshal and J.~Honorio.
\newblock Direct estimation of difference between structural equation models in
  high dimensions.
\newblock \emph{arXiv preprint arXiv:1906.12024}, 2019.

\bibitem[Gill et~al.(2014)Gill, Datta, and Datta]{Gilletal2014}
R.~Gill, S.~Datta, and S.~Datta.
\newblock dna: An {R} package for differential network analysis.
\newblock \emph{Bioinformation}, 10\penalty0 (4):\penalty0 233--234, 2014.

\bibitem[Goh et~al.(2007)Goh, Cusick, Valle, Childs, Vidal, and
  Barab{\'a}si]{goh2007}
K.-I. Goh, M.~E. Cusick, D.~Valle, B.~Childs, M.~Vidal, and A.-L. Barab{\'a}si.
\newblock The human disease network.
\newblock \emph{Proceedings of the National Academy of Sciences}, 104\penalty0
  (21):\penalty0 8685--8690, 2007.

\bibitem[Gomez-Ramirez and Wu(2014)]{gomez2014}
J.~Gomez-Ramirez and J.~Wu.
\newblock Network-based biomarkers in alzheimer’s disease: review and future
  directions.
\newblock \emph{Frontiers in aging neuroscience}, 6:\penalty0 12, 2014.

\bibitem[Guo et~al.(2011)Guo, Levina, Michailidis, and Zhu]{guo2011}
J.~Guo, E.~Levina, G.~Michailidis, and J.~Zhu.
\newblock Joint estimation of multiple graphical models.
\newblock \emph{Biometrika}, 98\penalty0 (1):\penalty0 1--15, 2011.

\bibitem[Hair et~al.(1998)Hair, Black, Babin, Anderson, and Tatham]{hair1998}
J.~F. Hair, W.~C. Black, B.~J. Babin, R.~E. Anderson, and R.~L. Tatham.
\newblock \emph{Multivariate data analysis}, volume~5.
\newblock Prentice hall Upper Saddle River, NJ, 1998.

\bibitem[Hao et~al.(2017)Hao, Sun, Liu, and Cheng]{hao2017}
B.~Hao, W.~W. Sun, Y.~Liu, and G.~Cheng.
\newblock Simultaneous clustering and estimation of heterogeneous graphical
  models.
\newblock \emph{The Journal of Machine Learning Research}, 18\penalty0
  (1):\penalty0 7981--8038, 2017.

\bibitem[He et~al.(2019)He, Cao, Zhang, Shen, Wang, and Deng]{he2019}
H.~He, S.~Cao, J.-g. Zhang, H.~Shen, Y.-P. Wang, and H.-w. Deng.
\newblock A statistical test for differential network analysis based on
  inference of gaussian graphical model.
\newblock \emph{Scientific reports}, 9\penalty0 (1):\penalty0 1--8, 2019.

\bibitem[Hotelling(1931)]{hotelling1931}
H.~Hotelling.
\newblock The generalization of student's ratio.
\newblock \emph{The Annals of Mathematical Statistics}, 2\penalty0
  (3):\penalty0 360--378, 1931.

\bibitem[Huang et~al.(2011)Huang, Ma, Li, and Zhang]{huang2011}
J.~Huang, S.~Ma, H.~Li, and C.-H. Zhang.
\newblock The sparse laplacian shrinkage estimator for high-dimensional
  regression.
\newblock \emph{Annals of statistics}, 39\penalty0 (4):\penalty0 2021, 2011.

\bibitem[Hussain and Harris(2006)]{hussain2006}
S.~P. Hussain and C.~C. Harris.
\newblock p53 biological network: at the crossroads of the cellular-stress
  response pathway and molecular carcinogenesis.
\newblock \emph{Journal of Nippon Medical School}, 73\penalty0 (2):\penalty0
  54--64, 2006.

\bibitem[Hyv{\"a}rinen(2005)]{hyvarinen2005}
A.~Hyv{\"a}rinen.
\newblock Estimation of non-normalized statistical models by score matching.
\newblock \emph{Journal of Machine Learning Research}, 6\penalty0
  (Apr):\penalty0 695--709, 2005.

\bibitem[Hyv{\"a}rinen(2007)]{hyvarinen2007}
A.~Hyv{\"a}rinen.
\newblock Some extensions of score matching.
\newblock \emph{Computational statistics \& data analysis}, 51\penalty0
  (5):\penalty0 2499--2512, 2007.

\bibitem[Ideker and Krogan(2012)]{ideker2012}
T.~Ideker and N.~J. Krogan.
\newblock Differential network biology.
\newblock \emph{Molecular systems biology}, 8\penalty0 (1), 2012.

\bibitem[Jacob et~al.(2012)Jacob, Neuvial, and Dudoit]{jacob2012}
L.~Jacob, P.~Neuvial, and S.~Dudoit.
\newblock More power via graph-structured tests for differential expression of
  gene networks.
\newblock \emph{The Annals of Applied Statistics}, 6\penalty0 (2):\penalty0
  561--600, 2012.

\bibitem[Jankov\'a and van~de Geer(2015)]{JankovavadeGeer2015}
J.~Jankov\'a and S.~van~de Geer.
\newblock Confidence intervals for high-dimensional inverse covariance
  estimation.
\newblock \emph{Electronic Journal of Statistics}, 9\penalty0 (1):\penalty0
  1205--1229, 2015.

\bibitem[Jankov\'a and van~de Geer(2017)]{JankovavandeGeer2017}
J.~Jankov\'a and S.~van~de Geer.
\newblock Honest confidence regions and optimality in high-dimensional
  precision matrix estimation.
\newblock \emph{TEST}, 26\penalty0 (1):\penalty0 143--162, 2017.

\bibitem[Junker and Schreiber(2008)]{junker2008}
B.~H. Junker and F.~Schreiber.
\newblock \emph{Analysis of biological networks}, volume~2.
\newblock Wiley Online Library, 2008.

\bibitem[Kerr and Churchill(2001)]{kerr2001}
M.~K. Kerr and G.~A. Churchill.
\newblock Statistical design and the analysis of gene expression microarray
  data.
\newblock \emph{Genetics Research}, 77\penalty0 (2):\penalty0 123--128, 2001.

\bibitem[Khare et~al.(2015)Khare, Oh, and Rajaratnam]{khare2015}
K.~Khare, S.-Y. Oh, and B.~Rajaratnam.
\newblock A convex pseudolikelihood framework for high dimensional partial
  correlation estimation with convergence guarantees.
\newblock \emph{Journal of the Royal Statistical Society: Series B (Statistical
  Methodology)}, 77\penalty0 (4):\penalty0 803--825, 2015.

\bibitem[Khatri and Rao(1976)]{khatri1976}
C.~Khatri and C.~R. Rao.
\newblock Characterizations of multivariate normality. i. through independence
  of some statistics.
\newblock \emph{Journal of Multivariate Analysis}, 6\penalty0 (1):\penalty0
  81--94, 1976.

\bibitem[Khatri et~al.(2012)Khatri, Sirota, and Butte]{khatri2012}
P.~Khatri, M.~Sirota, and A.~J. Butte.
\newblock Ten years of pathway analysis: current approaches and outstanding
  challenges.
\newblock \emph{PLoS computational biology}, 8\penalty0 (2):\penalty0 e1002375,
  2012.

\bibitem[Kim et~al.(2019)Kim, Liu, and Kolar]{kim2019}
B.~Kim, S.~Liu, and M.~Kolar.
\newblock Two-sample inference for high-dimensional markov networks.
\newblock \emph{arXiv preprint arXiv:1905.00466}, 2019.

\bibitem[Kitano(2004)]{Kitano2004}
H.~Kitano.
\newblock Biological robustness.
\newblock \emph{Nature Reviews Genetics}, 5:\penalty0 826--837, 2004.

\bibitem[Koller and Friedman(2009)]{koller2009}
D.~Koller and N.~Friedman.
\newblock \emph{Probabilistic graphical models: principles and techniques}.
\newblock MIT press, 2009.

\bibitem[Krogan et~al.(2006)Krogan, Cagney, Yu, Zhong, Guo, Ignatchenko, Li,
  Pu, Datta, and Tikuisis]{krogan2006}
N.~J. Krogan, G.~Cagney, H.~Yu, G.~Zhong, X.~Guo, A.~Ignatchenko, J.~Li, S.~Pu,
  N.~Datta, and A.~P. Tikuisis.
\newblock Global landscape of protein complexes in the yeast saccharomyces
  cerevisiae.
\newblock \emph{Nature}, 440\penalty0 (7084):\penalty0 637, 2006.

\bibitem[Landt et~al.(2012)Landt, Marinov, Kundaje, Kheradpour, Pauli,
  Batzoglou, Bernstein, Bickel, Brown, and Cayting]{landt2012}
S.~G. Landt, G.~K. Marinov, A.~Kundaje, P.~Kheradpour, F.~Pauli, S.~Batzoglou,
  B.~E. Bernstein, P.~Bickel, J.~B. Brown, and P.~Cayting.
\newblock Chip-seq guidelines and practices of the encode and modencode
  consortia.
\newblock \emph{Genome research}, 22\penalty0 (9):\penalty0 1813--1831, 2012.

\bibitem[Langfelder and Horvath(2008)]{langfelder2008}
P.~Langfelder and S.~Horvath.
\newblock Wgcna: an r package for weighted correlation network analysis.
\newblock \emph{BMC bioinformatics}, 9\penalty0 (1):\penalty0 559, 2008.

\bibitem[Lauritzen(1996)]{lauritzen1996}
S.~L. Lauritzen.
\newblock \emph{Graphical models}, volume~17.
\newblock Clarendon Press, 1996.

\bibitem[Lee et~al.(2016)Lee, Li, and Zhao]{lee2016}
K.-Y. Lee, B.~Li, and H.~Zhao.
\newblock On an additive partial correlation operator and nonparametric
  estimation of graphical models.
\newblock \emph{Biometrika}, 103\penalty0 (3):\penalty0 513--530, 2016.

\bibitem[Leek et~al.(2010)Leek, Scharpf, Bravo, Simcha, Langmead, Johnson,
  Geman, Baggerly, and Irizarry]{leek2010}
J.~T. Leek, R.~B. Scharpf, H.~C. Bravo, D.~Simcha, B.~Langmead, W.~E. Johnson,
  D.~Geman, K.~Baggerly, and R.~A. Irizarry.
\newblock Tackling the widespread and critical impact of batch effects in
  high-throughput data.
\newblock \emph{Nature Reviews Genetics}, 11\penalty0 (10):\penalty0 733, 2010.

\bibitem[Li and Chen(2012)]{li2012}
J.~Li and S.~X. Chen.
\newblock Two sample tests for high-dimensional covariance matrices.
\newblock \emph{The Annals of Statistics}, 40\penalty0 (2):\penalty0 908--940,
  2012.

\bibitem[Lin et~al.(2016)Lin, Drton, and Shojaie]{lin2016}
L.~Lin, M.~Drton, and A.~Shojaie.
\newblock Estimation of high-dimensional graphical models using regularized
  score matching.
\newblock \emph{Electronic journal of statistics}, 10\penalty0 (1):\penalty0
  806, 2016.

\bibitem[Liu et~al.(2009)Liu, Lafferty, and Wasserman]{liu2009}
H.~Liu, J.~Lafferty, and L.~Wasserman.
\newblock The nonparanormal: Semiparametric estimation of high dimensional
  undirected graphs.
\newblock \emph{Journal of Machine Learning Research}, 10\penalty0
  (Oct):\penalty0 2295--2328, 2009.

\bibitem[Liu et~al.(2012)Liu, Han, Yuan, Lafferty, and Wasserman]{liu2012}
H.~Liu, F.~Han, M.~Yuan, J.~Lafferty, and L.~Wasserman.
\newblock High-dimensional semiparametric gaussian copula graphical models.
\newblock \emph{The Annals of Statistics}, 40\penalty0 (4):\penalty0
  2293--2326, 2012.

\bibitem[Liu(2016)]{liu2016}
Z.-P. Liu.
\newblock Identifying network-based biomarkers of complex diseases from
  high-throughput data.
\newblock \emph{Biomarkers in Medicine}, 10\penalty0 (6):\penalty0 633--650,
  2016.

\bibitem[Luscombe et~al.(2004)Luscombe, Babu, Yu, Snyder, Teichmann, and
  Gerstein]{luscombe2004}
N.~M. Luscombe, M.~M. Babu, H.~Yu, M.~Snyder, S.~A. Teichmann, and M.~Gerstein.
\newblock Genomic analysis of regulatory network dynamics reveals large
  topological changes.
\newblock \emph{Nature}, 431\penalty0 (7006):\penalty0 308, 2004.

\bibitem[Ma et~al.(2014)Ma, Xin, Feldmann, and Wang]{ma2014}
C.~Ma, M.~Xin, K.~A. Feldmann, and X.~Wang.
\newblock Machine learning--based differential network analysis: A study of
  stress-responsive transcriptomes in arabidopsis.
\newblock \emph{The Plant Cell}, 26\penalty0 (2):\penalty0 520--537, 2014.

\bibitem[Ma and Michailidis(2016)]{ma2016}
J.~Ma and G.~Michailidis.
\newblock Joint structural estimation of multiple graphical models.
\newblock \emph{The Journal of Machine Learning Research}, 17\penalty0
  (1):\penalty0 5777--5824, 2016.

\bibitem[Ma et~al.(2016)Ma, Shojaie, and Michailidis]{ma2016netgsa}
J.~Ma, A.~Shojaie, and G.~Michailidis.
\newblock Network-based pathway enrichment analysis with incomplete network
  information.
\newblock \emph{Bioinformatics}, 32\penalty0 (20):\penalty0 3165--3174, 2016.

\bibitem[Ma et~al.(2019{\natexlab{a}})Ma, Karnovsky, Afshinnia, Wigginton,
  Rader, Natarajan, Sharma, Porter, Rahman, He, Hamm, Shafi, Gipson, Gadegbeku,
  Feldman, Michailidis, Pennathur, and the CPROBE~study
  investigators]{ma2019diff}
J.~Ma, A.~Karnovsky, F.~Afshinnia, J.~Wigginton, D.~J. Rader, L.~Natarajan,
  K.~Sharma, A.~C. Porter, M.~Rahman, J.~He, L.~Hamm, T.~Shafi, D.~Gipson,
  C.~Gadegbeku, H.~Feldman, G.~Michailidis, t.~C. Pennathur, Subramaniam, and
  the CPROBE~study investigators.
\newblock {Differential network enrichment analysis reveals novel lipid
  pathways in chronic kidney disease}.
\newblock \emph{Bioinformatics}, 35\penalty0 (18):\penalty0 3441--3452,
  2019{\natexlab{a}}.

\bibitem[Ma et~al.(2019{\natexlab{b}})Ma, Shojaie, and Michailidis]{ma2019}
J.~Ma, A.~Shojaie, and G.~Michailidis.
\newblock A comparative study of topology-based pathway enrichment analysis
  methods.
\newblock \emph{BMC bioinformatics}, 20\penalty0 (1):\penalty0 546,
  2019{\natexlab{b}}.

\bibitem[Manzour et~al.(2019)Manzour, K{\"u}{\c{c}}{\"u}kyavuz, and
  Shojaie]{manzour2019}
H.~Manzour, S.~K{\"u}{\c{c}}{\"u}kyavuz, and A.~Shojaie.
\newblock Integer programming for learning directed acyclic graphs from
  continuous data.
\newblock \emph{arXiv preprint arXiv:1904.10574}, 2019.

\bibitem[Margolin et~al.(2006)Margolin, Nemenman, Basso, Wiggins, Stolovitzky,
  Dalla~Favera, and Califano]{margolin2006}
A.~A. Margolin, I.~Nemenman, K.~Basso, C.~Wiggins, G.~Stolovitzky,
  R.~Dalla~Favera, and A.~Califano.
\newblock Aracne: an algorithm for the reconstruction of gene regulatory
  networks in a mammalian cellular context.
\newblock In \emph{BMC bioinformatics}, volume~7, page~S7. BioMed Central,
  2006.

\bibitem[Markowetz and Spang(2007)]{markowetz2007}
F.~Markowetz and R.~Spang.
\newblock Inferring cellular networks--a review.
\newblock \emph{BMC bioinformatics}, 8\penalty0 (6):\penalty0 S5, 2007.

\bibitem[Massa et~al.(2010)Massa, Chiogna, and Romualdi]{massa2010}
M.~S. Massa, M.~Chiogna, and C.~Romualdi.
\newblock Gene set analysis exploiting the topology of a pathway.
\newblock \emph{BMC Systems Biology}, 4\penalty0 (1):\penalty0 121, 2010.

\bibitem[McKenzie et~al.(2016)McKenzie, Katsyv, Song, Wang, and
  Zhang]{mckenzie2016}
A.~T. McKenzie, I.~Katsyv, W.-M. Song, M.~Wang, and B.~Zhang.
\newblock Dgca: a comprehensive r package for differential gene correlation
  analysis.
\newblock \emph{BMC systems biology}, 10\penalty0 (1):\penalty0 106, 2016.

\bibitem[Meinshausen and B{\"u}hlmann(2006)]{meinshausen2006}
N.~Meinshausen and P.~B{\"u}hlmann.
\newblock High-dimensional graphs and variable selection with the lasso.
\newblock \emph{The Annals of Statistics}, 34\penalty0 (3):\penalty0
  1436--1462, 2006.

\bibitem[Na et~al.(2019)Na, Kolar, and Koyejo]{na2019estimating}
S.~Na, M.~Kolar, and O.~Koyejo.
\newblock Estimating differential latent variable graphical models with
  applications to brain connectivity.
\newblock \emph{arXiv preprint arXiv:1909.05892}, 2019.

\bibitem[Pearl(2009)]{pearl2009}
J.~Pearl.
\newblock \emph{Causality}.
\newblock Cambridge university press, 2009.

\bibitem[Peng et~al.(2009)Peng, Wang, Zhou, and Zhu]{peng2009}
J.~Peng, P.~Wang, N.~Zhou, and J.~Zhu.
\newblock Partial correlation estimation by joint sparse regression models.
\newblock \emph{Journal of the American Statistical Association}, 104\penalty0
  (486):\penalty0 735--746, 2009.

\bibitem[Peters and B{\"u}hlmann(2013)]{peters2013}
J.~Peters and P.~B{\"u}hlmann.
\newblock Identifiability of gaussian structural equation models with equal
  error variances.
\newblock \emph{Biometrika}, 101\penalty0 (1):\penalty0 219--228, 2013.

\bibitem[Peterson et~al.(2015)Peterson, Stingo, and Vannucci]{peterson2015}
C.~Peterson, F.~C. Stingo, and M.~Vannucci.
\newblock Bayesian inference of multiple gaussian graphical models.
\newblock \emph{Journal of the American Statistical Association}, 110\penalty0
  (509):\penalty0 159--174, 2015.

\bibitem[Pourahmadi(2013)]{pourahmadi2013}
M.~Pourahmadi.
\newblock \emph{High-dimensional covariance estimation: with high-dimensional
  data}, volume 882.
\newblock John Wiley \& Sons, 2013.

\bibitem[Ravikumar et~al.(2010)Ravikumar, Wainwright, Lafferty,
  et~al.]{ravikumar2010}
P.~Ravikumar, M.~J. Wainwright, J.~D. Lafferty, et~al.
\newblock High-dimensional ising model selection using ℓ1-regularized
  logistic regression.
\newblock \emph{The Annals of Statistics}, 38\penalty0 (3):\penalty0
  1287--1319, 2010.

\bibitem[Ren et~al.(2015)Ren, Sun, Zhang, and Zhou]{Renetal2015}
Z.~Ren, T.~Sun, C.-H. Zhang, and H.~H. Zhou.
\newblock Asymptotic normality and optimalities in estimation of large
  {Gaussian} graphical models.
\newblock \emph{The Annals of Statistics}, 43\penalty0 (3):\penalty0 991--1026,
  2015.

\bibitem[Rothman et~al.(2008)Rothman, Bickel, Levina, and Zhu]{rothman2008}
A.~J. Rothman, P.~J. Bickel, E.~Levina, and J.~Zhu.
\newblock Sparse permutation invariant covariance estimation.
\newblock \emph{Electronic Journal of Statistics}, 2:\penalty0 494--515, 2008.

\bibitem[Saegusa and Shojaie(2016)]{saegusa2016}
T.~Saegusa and A.~Shojaie.
\newblock Joint estimation of precision matrices in heterogeneous populations.
\newblock \emph{Electronic journal of statistics}, 10\penalty0 (1):\penalty0
  1341, 2016.

\bibitem[Sas et~al.(2018)Sas, Lin, Rajendiran, Soni, Nair, Hinder, Jagadish,
  Gardner, Abcouwer, Brosius, et~al.]{sas2018}
K.~M. Sas, J.~Lin, T.~M. Rajendiran, T.~Soni, V.~Nair, L.~M. Hinder, H.~V.
  Jagadish, T.~W. Gardner, S.~F. Abcouwer, F.~C. Brosius, et~al.
\newblock Shared and distinct lipid-lipid interactions in plasma and affected
  tissues in a diabetic mouse model.
\newblock \emph{Journal of lipid research}, 59\penalty0 (2):\penalty0 173--183,
  2018.

\bibitem[Schmidt et~al.(2010)Schmidt, Wilson, Ballester, Schwalie, Brown,
  Marshall, Kutter, Watt, Martinez-Jimenez, and Mackay]{schmidt2010}
D.~Schmidt, M.~D. Wilson, B.~Ballester, P.~C. Schwalie, G.~D. Brown,
  A.~Marshall, C.~Kutter, S.~Watt, C.~P. Martinez-Jimenez, and S.~Mackay.
\newblock Five-vertebrate chip-seq reveals the evolutionary dynamics of
  transcription factor binding.
\newblock \emph{Science}, 328\penalty0 (5981):\penalty0 1036--1040, 2010.

\bibitem[Schott(2007)]{schott2007}
J.~R. Schott.
\newblock A test for the equality of covariance matrices when the dimension is
  large relative to the sample sizes.
\newblock \emph{Computational Statistics \& Data Analysis}, 51\penalty0
  (12):\penalty0 6535--6542, 2007.

\bibitem[Shojaie and Michailidis(2009)]{shojaie2009}
A.~Shojaie and G.~Michailidis.
\newblock Analysis of gene sets based on the underlying regulatory network.
\newblock \emph{Journal of Computational Biology}, 16\penalty0 (3):\penalty0
  407--426, 2009.

\bibitem[Shojaie and Michailidis(2010)]{shojaie2010}
A.~Shojaie and G.~Michailidis.
\newblock Network enrichment analysis in complex experiments.
\newblock \emph{Statistical applications in genetics and molecular biology},
  9\penalty0 (1), 2010.

\bibitem[Shojaie and Sedaghat(2017)]{shojaie2017}
A.~Shojaie and N.~Sedaghat.
\newblock How different are estimated genetic networks of cancer subtypes?
\newblock In \emph{Big and Complex Data Analysis}, pages 159--192. Springer,
  2017.

\bibitem[Simon and Tibshirani(2012)]{simon2012standardization}
N.~Simon and R.~Tibshirani.
\newblock Standardization and the group lasso penalty.
\newblock \emph{Statistica Sinica}, 22:\penalty0 983--1001, 2012.

\bibitem[Smith et~al.(1962)Smith, Gnanadesikan, and Hughes]{smith1962}
H.~Smith, R.~Gnanadesikan, and J.~Hughes.
\newblock Multivariate analysis of variance (manova).
\newblock \emph{Biometrics}, 18\penalty0 (1):\penalty0 22--41, 1962.

\bibitem[Srivastava and Yanagihara(2010)]{srivastava2010}
M.~S. Srivastava and H.~Yanagihara.
\newblock Testing the equality of several covariance matrices with fewer
  observations than the dimension.
\newblock \emph{Journal of Multivariate Analysis}, 101\penalty0 (6):\penalty0
  1319--1329, 2010.

\bibitem[Stelzl et~al.(2005)Stelzl, Worm, Lalowski, Haenig, Brembeck, Goehler,
  Stroedicke, Zenkner, Schoenherr, Koeppen, et~al.]{stelzl2005}
U.~Stelzl, U.~Worm, M.~Lalowski, C.~Haenig, F.~H. Brembeck, H.~Goehler,
  M.~Stroedicke, M.~Zenkner, A.~Schoenherr, S.~Koeppen, et~al.
\newblock A human protein-protein interaction network: a resource for
  annotating the proteome.
\newblock \emph{Cell}, 122\penalty0 (6):\penalty0 957--968, 2005.

\bibitem[Subramanian et~al.(2005)Subramanian, Tamayo, Mootha, Mukherjee, Ebert,
  Gillette, Paulovich, Pomeroy, Golub, and Lander]{subramanian2005}
A.~Subramanian, P.~Tamayo, V.~K. Mootha, S.~Mukherjee, B.~L. Ebert, M.~A.
  Gillette, A.~Paulovich, S.~L. Pomeroy, T.~R. Golub, and E.~S. Lander.
\newblock Gene set enrichment analysis: a knowledge-based approach for
  interpreting genome-wide expression profiles.
\newblock \emph{Proceedings of the National Academy of Sciences}, 102\penalty0
  (43):\penalty0 15545--15550, 2005.

\bibitem[Suggala et~al.(2017)Suggala, Kolar, and
  Ravikumar]{suggala2017expxorcist}
A.~Suggala, M.~Kolar, and P.~K. Ravikumar.
\newblock The expxorcist: nonparametric graphical models via conditional
  exponential densities.
\newblock In \emph{{Advances in Neural Information Processing Systems}}, pages
  4446--4456, 2017.

\bibitem[Sun et~al.(2015)Sun, Kolar, and Xu]{sun2015learning}
S.~Sun, M.~Kolar, and J.~Xu.
\newblock Learning structured densities via infinite dimensional exponential
  families.
\newblock In \emph{{Advances in Neural Information Processing Systems}}, pages
  2287--2295, 2015.

\bibitem[Sz{\'e}kely and Rizzo(2013)]{szekely2013}
G.~J. Sz{\'e}kely and M.~L. Rizzo.
\newblock Energy statistics: A class of statistics based on distances.
\newblock \emph{Journal of statistical planning and inference}, 143\penalty0
  (8):\penalty0 1249--1272, 2013.

\bibitem[Tarassov et~al.(2008)Tarassov, Messier, Landry, Radinovic, Molina,
  Shames, Malitskaya, Vogel, Bussey, and Michnick]{tarassov2008}
K.~Tarassov, V.~Messier, C.~R. Landry, S.~Radinovic, M.~M.~S. Molina,
  I.~Shames, Y.~Malitskaya, J.~Vogel, H.~Bussey, and S.~W. Michnick.
\newblock An in vivo map of the yeast protein interactome.
\newblock \emph{Science}, 320\penalty0 (5882):\penalty0 1465--1470, 2008.

\bibitem[Taylor et~al.(2009)Taylor, Linding, Warde-Farley, Liu, Pesquita,
  Faria, Bull, Pawson, Morris, and Wrana]{taylor2009}
I.~W. Taylor, R.~Linding, D.~Warde-Farley, Y.~Liu, C.~Pesquita, D.~Faria,
  S.~Bull, T.~Pawson, Q.~Morris, and J.~L. Wrana.
\newblock Dynamic modularity in protein interaction networks predicts breast
  cancer outcome.
\newblock \emph{Nature biotechnology}, 27\penalty0 (2):\penalty0 199, 2009.

\bibitem[Tian et~al.(2005)Tian, Greenberg, Kong, Altschuler, Kohane, and
  Park]{tian2005}
L.~Tian, S.~A. Greenberg, S.~W. Kong, J.~Altschuler, I.~S. Kohane, and P.~J.
  Park.
\newblock Discovering statistically significant pathways in expression
  profiling studies.
\newblock \emph{Proceedings of the National Academy of Sciences}, 102\penalty0
  (38):\penalty0 13544--13549, 2005.

\bibitem[Tibshirani(1996)]{tibshirani1996}
R.~Tibshirani.
\newblock Regression shrinkage and selection via the lasso.
\newblock \emph{Journal of the Royal Statistical Society: Series B
  (Methodological)}, 58\penalty0 (1):\penalty0 267--288, 1996.

\bibitem[Tibshirani et~al.(2005)Tibshirani, Saunders, Rosset, Zhu, and
  Knight]{tibshirani2005}
R.~Tibshirani, M.~Saunders, S.~Rosset, J.~Zhu, and K.~Knight.
\newblock Sparsity and smoothness via the fused lasso.
\newblock \emph{Journal of the Royal Statistical Society: Series B (Statistical
  Methodology)}, 67\penalty0 (1):\penalty0 91--108, 2005.

\bibitem[Troy et~al.(2016)Troy, Hollams, Holt, and Bosco]{troy2016}
N.~M. Troy, E.~M. Hollams, P.~G. Holt, and A.~Bosco.
\newblock Differential gene network analysis for the identification of
  asthma-associated therapeutic targets in allergen-specific t-helper memory
  responses.
\newblock \emph{BMC medical genomics}, 9\penalty0 (1):\penalty0 9, 2016.

\bibitem[Tryputsen et~al.(2015)Tryputsen, DiBernardo, Samtani, Novak, Narayan,
  Raghavan, and Initiative]{tryputsen2015}
V.~Tryputsen, A.~DiBernardo, M.~Samtani, G.~P. Novak, V.~A. Narayan,
  N.~Raghavan, and A.~D.~N. Initiative.
\newblock Optimizing regions-of-interest composites for capturing treatment
  effects on brain amyloid in clinical trials.
\newblock \emph{Journal of Alzheimer's Disease}, 43\penalty0 (3):\penalty0
  809--821, 2015.

\bibitem[Voorman et~al.(2013)Voorman, Shojaie, and Witten]{voorman2013}
A.~Voorman, A.~Shojaie, and D.~Witten.
\newblock Graph estimation with joint additive models.
\newblock \emph{Biometrika}, 101\penalty0 (1):\penalty0 85--101, 2013.

\bibitem[Wainwright et~al.(2008)Wainwright, Jordan, et~al.]{wainwright2008}
M.~J. Wainwright, M.~I. Jordan, et~al.
\newblock Graphical models, exponential families, and variational inference.
\newblock \emph{Foundations and Trends{\textregistered} in Machine Learning},
  1\penalty0 (1--2):\penalty0 1--305, 2008.

\bibitem[Wang et~al.(2012)]{wang2012}
H.~Wang et~al.
\newblock Bayesian graphical lasso models and efficient posterior computation.
\newblock \emph{Bayesian Analysis}, 7\penalty0 (4):\penalty0 867--886, 2012.

\bibitem[Wang et~al.(2006)Wang, Joshi, Zhang, Xu, and Chen]{wang2006}
Y.~Wang, T.~Joshi, X.-S. Zhang, D.~Xu, and L.~Chen.
\newblock Inferring gene regulatory networks from multiple microarray datasets.
\newblock \emph{Bioinformatics}, 22\penalty0 (19):\penalty0 2413--2420, 2006.

\bibitem[Wang et~al.(2018)Wang, Squires, Belyaeva, and Uhler]{wang2018}
Y.~Wang, C.~Squires, A.~Belyaeva, and C.~Uhler.
\newblock Direct estimation of differences in causal graphs.
\newblock In \emph{Advances in Neural Information Processing Systems}, pages
  3770--3781, 2018.

\bibitem[West et~al.(2012)West, Bianconi, Severini, and Teschendorff]{west2012}
J.~West, G.~Bianconi, S.~Severini, and A.~E. Teschendorff.
\newblock Differential network entropy reveals cancer system hallmarks.
\newblock \emph{Scientific reports}, 2:\penalty0 802, 2012.

\bibitem[Wu and Li(2015)]{wu2015}
T.-L. Wu and P.~Li.
\newblock Tests for high-dimensional covariance matrices using random matrix
  projection.
\newblock \emph{arXiv preprint arXiv:1511.01611}, 2015.

\bibitem[Xia and Li(2017)]{XiaLi2017}
Y.~Xia and L.~Li.
\newblock Hypothesis testing of matrix graph model with application to brain
  connectivity analysis.
\newblock \emph{Biometrics}, 73\penalty0 (3):\penalty0 780--791, 2017.

\bibitem[Xia et~al.(2015)Xia, Cai, and Cai]{Xiaetal2015}
Y.~Xia, T.~Cai, and T.~T. Cai.
\newblock Testing differential networks with applications to detecting
  gene-by-gene interactions.
\newblock \emph{Biometrika}, 102\penalty0 (2):\penalty0 247--266, 2015.

\bibitem[Xue et~al.(2012)Xue, Zou, et~al.]{xue2012}
L.~Xue, H.~Zou, et~al.
\newblock Regularized rank-based estimation of high-dimensional nonparanormal
  graphical models.
\newblock \emph{The Annals of Statistics}, 40\penalty0 (5):\penalty0
  2541--2571, 2012.

\bibitem[Yamanishi et~al.(2004)Yamanishi, Vert, and Kanehisa]{yamanishi2004}
Y.~Yamanishi, J.-P. Vert, and M.~Kanehisa.
\newblock Protein network inference from multiple genomic data: a supervised
  approach.
\newblock \emph{Bioinformatics}, 20\penalty0 (suppl\_1):\penalty0 i363--i370,
  2004.

\bibitem[Yang et~al.(2012)Yang, Allen, Liu, and Ravikumar]{yang2012}
E.~Yang, G.~Allen, Z.~Liu, and P.~K. Ravikumar.
\newblock Graphical models via generalized linear models.
\newblock In \emph{Advances in Neural Information Processing Systems}, pages
  1358--1366, 2012.

\bibitem[Yang et~al.(2013)Yang, Ravikumar, Allen, and Liu]{yang2013}
E.~Yang, P.~K. Ravikumar, G.~I. Allen, and Z.~Liu.
\newblock On poisson graphical models.
\newblock In \emph{Advances in Neural Information Processing Systems}, pages
  1718--1726, 2013.

\bibitem[Yang et~al.(2014)Yang, Baker, Ravikumar, Allen, and Liu]{yang2014}
E.~Yang, Y.~Baker, P.~Ravikumar, G.~Allen, and Z.~Liu.
\newblock Mixed graphical models via exponential families.
\newblock In \emph{Artificial Intelligence and Statistics}, pages 1042--1050,
  2014.

\bibitem[Yu et~al.(2019{\natexlab{a}})Yu, Gupta, and Kolar]{yu2019simultaneous}
M.~Yu, V.~Gupta, and M.~Kolar.
\newblock Simultaneous inference for pairwise graphical models with generalized
  score matching.
\newblock \emph{arXiv preprint arXiv:1905.06261}, 2019{\natexlab{a}}.

\bibitem[Yu et~al.(2018)Yu, Drton, and Shojaie]{yu2018}
S.~Yu, M.~Drton, and A.~Shojaie.
\newblock Graphical models for non-negative data using generalized score
  matching.
\newblock In \emph{International Conference on Artificial Intelligence and
  Statistics}, pages 1781--1790, 2018.

\bibitem[Yu et~al.(2019{\natexlab{b}})Yu, Drton, and Shojaie]{yu2019}
S.~Yu, M.~Drton, and A.~Shojaie.
\newblock Generalized score matching for non-negative data.
\newblock \emph{Journal of Machine Learning Research}, 20\penalty0
  (76):\penalty0 1--70, 2019{\natexlab{b}}.

\bibitem[Yuan et~al.(2017)Yuan, Xi, Chen, and Deng]{yuan2017}
H.~Yuan, R.~Xi, C.~Chen, and M.~Deng.
\newblock Differential network analysis via lasso penalized d-trace loss.
\newblock \emph{Biometrika}, 104\penalty0 (4):\penalty0 755--770, 2017.

\bibitem[Yuan and Lin(2006)]{yuan2006}
M.~Yuan and Y.~Lin.
\newblock Model selection and estimation in regression with grouped variables.
\newblock \emph{Journal of the Royal Statistical Society: Series B (Statistical
  Methodology)}, 68\penalty0 (1):\penalty0 49--67, 2006.

\bibitem[Yuan and Lin(2007)]{yuan2007}
M.~Yuan and Y.~Lin.
\newblock Model selection and estimation in the gaussian graphical model.
\newblock \emph{Biometrika}, 94\penalty0 (1):\penalty0 19--35, 2007.

\bibitem[Zhang and Zou(2014)]{zhang2014}
T.~Zhang and H.~Zou.
\newblock Sparse precision matrix estimation via lasso penalized d-trace loss.
\newblock \emph{Biometrika}, 101\penalty0 (1):\penalty0 103--120, 2014.

\bibitem[Zhang et~al.(2016)Zhang, Ou-Yang, Zhao, and Yan]{zhang2016}
X.-F. Zhang, L.~Ou-Yang, X.-M. Zhao, and H.~Yan.
\newblock Differential network analysis from cross-platform gene expression
  data.
\newblock \emph{Scientific reports}, 6:\penalty0 34112, 2016.

\bibitem[Zhao et~al.(2019)Zhao, Ottinger, Peck, Mac~Donald, and
  Shojaie]{zhao2019}
S.~Zhao, S.~Ottinger, S.~Peck, C.~Mac~Donald, and A.~Shojaie.
\newblock Network differential connectivity analysis.
\newblock \emph{arXiv preprint arXiv:1909.13464}, 2019.

\bibitem[Zhao et~al.(2014)Zhao, Cai, and Li]{zhao2014}
S.~D. Zhao, T.~T. Cai, and H.~Li.
\newblock Direct estimation of differential networks.
\newblock \emph{Biometrika}, 101\penalty0 (2):\penalty0 253--268, 2014.

\bibitem[Zhong et~al.(2009)Zhong, Simonis, Li, Charloteaux, Heuze, Klitgord,
  Tam, Yu, Venkatesan, Mou, et~al.]{zhong2009}
Q.~Zhong, N.~Simonis, Q.-R. Li, B.~Charloteaux, F.~Heuze, N.~Klitgord, S.~Tam,
  H.~Yu, K.~Venkatesan, D.~Mou, et~al.
\newblock Edgetic perturbation models of human inherited disorders.
\newblock \emph{Molecular systems biology}, 5\penalty0 (1), 2009.

\bibitem[Zhu et~al.(2017)Zhu, Lei, Devlin, and Roeder]{zhu2017}
L.~Zhu, J.~Lei, B.~Devlin, and K.~Roeder.
\newblock Testing high-dimensional covariance matrices, with application to
  detecting schizophrenia risk genes.
\newblock \emph{Annals of Applied Statistics}, 11\penalty0 (3):\penalty0 1810,
  2017.

\bibitem[Zhu et~al.(2014)Zhu, Shen, and Pan]{zhu2014}
Y.~Zhu, X.~Shen, and W.~Pan.
\newblock Structural pursuit over multiple undirected graphs.
\newblock \emph{Journal of the American Statistical Association}, 109\penalty0
  (508):\penalty0 1683--1696, 2014.

\end{thebibliography}

\end{document}